\documentclass[10pt]{article}
\title{Mediation analysis in longitudinal intervention studies with an ordinal treatment-dependent confounder}
\usepackage{authblk}
\author[1]{Mikko Valtanen}
\author[2]{Tommi Härkänen}
\author[3]{Matti Uusitupa}
\author[2,4]{Jaakko Tuomilehto}
\author[2,3]{Jaana Lindström}
\author[5]{Kari Auranen}

\affil[1]{\footnotesize Department of Mathematics and Statistics, 20014 University of Turku, Finland}
\affil[2]{\footnotesize Public Health, Finnish Institute for Health and Welfare, PO Box 30, FI-00271 Helsinki, Finland}
\affil[3]{\footnotesize Institute of Public Health and Clinical Nutrition, University of Eastern Finland, PO Box 1627, 70211 Kuopio}
\affil[4]{\footnotesize Department of Public Health, University of Helsinki, Tukholmankatu 8B, 00014 Helsinki, Finland}
\affil[5]{\footnotesize Department of Clinical Medicine, 20014 University of Turku, Finland}

\usepackage[utf8]{inputenc}
\usepackage[none]{hyphenat}
\usepackage{amsmath}
\usepackage{amssymb}
\usepackage{graphicx}
\usepackage{nicefrac}
\usepackage{mathtools}
\usepackage{verbatim}
\usepackage{multirow}
\usepackage{hhline}
\usepackage{makecell}
\usepackage{natbib}
\usepackage{booktabs}
\newcommand{\ind}{\perp\!\!\!\!\perp}
\newcommand{\dd}[1]{\,\text{d}#1}
\usepackage[a4paper, margin=1.5in]{geometry}  
\bibpunct{(}{)}{;}{a}{,}{,}
\date{}
\begin{document}

\maketitle
\begin{abstract}
In interventional health studies, causal mediation analysis can be employed to investigate mechanisms through which the intervention affects the targeted health outcome.
Identifying direct and indirect (i.e. mediated) effects from empirical data become complicated, however, when the mediator-outcome association is confounded by a variable itself affected by the treatment. 
Here, we investigate identification of mediational effects under such post-treatment confounding in a setting with a longitudinal mediator, time-to-event outcome and a trichotomous ordinal treatment-dependent confounder. 
If the intervention always affects the treatment-dependent confounder only in one direction (monotonicity), we show that the mediational effects are identified up to a stratum-specific sensitivity parameter and derive their empirical non-parametric expressions. 
The feasibility of the monotonicity assumption can be assessed using empirical data, based on restrictions on the marginal distributions of counterfactuals of the treatment-dependent confounder.
We avoid pitfalls related to post-treatment conditioning by treating the mediator as a functional entity and defining the time-to-event outcome as a restricted disease-free time. 
In an empirical analysis, we use data from the Finnish Diabetes Prevention Study to assess the extent to which the effect of a lifestyle intervention on avoiding type 2 diabetes is mediated through weight reduction in a high-risk population, with other health-related changes acting as treatment-dependent confounders.
\end{abstract}

\section{Introduction}
Lifestyle choices play an important role in the prevention of type 2 diabetes (T2D).
As T2D causes major health and economic burdens globally, lifestyle interventions to reduce its incidence in high-risk populations are an area of active research \citep{Tuomilehto2023,Uusitupa2019,Barry2017,Haw2017}.
Studies of such interventions typically follow a cohort of individuals over time and aim at collecting information on biomarkers and health status evaluations at several follow-up visits \citep{finger,Knowler2002,cardia}. 
In countries with extensive health registers, data gathered at follow-up visits can be augmented with time-to-event outcomes retrieved from the registers.

Intervention studies are often based on experimental designs where study participants are assigned to treatment and control groups and the target of inference is the causal effect of an intervention on a specific health response.
A more elaborate question involves understanding the extent to which the treatment effect is mediated through an intermediate variable.
For example, potential mediating mechanisms of liraglutide treatment on cardiovascular and chronic kidney disease through changes in clinical biomarkers have been previously investigated in people with diabetes \citep{Buse2020, Mann2021}.

Mediating mechanisms can be addressed within the causal mediation analysis framework, where the total effect of treatment is decomposed into direct and indirect (i.e. mediated) effects \citep{Robins1992, Pearl2001}. 
The estimands of these causal effects can be defined in terms of counterfactuals, i.e. expected outcomes in hypothetical scenarios where the treatment and the mediator are intervened upon to set them at certain values \citep{Rubin1974}.
Under non-trivial conditional independence assumptions, the estimands can be non-parametrically identified from empirical data \citep{Pearl2014}. 
Studies using a lifestyle intervention as the treatment and a clinical risk factor as the mediator, however, are especially prone to failing the identifiability assumptions, because the intervention often induces behavioural changes that affect the response through both the intended mediator and other mechanisms.
The behavioural change then acts as a treatment-dependent confounder, i.e. a variable that confounds the mediator--outcome relationship and is itself affected by the treatment.
In such situations the standard independence assumptions do not suffice to identify natural mediational effects non-parametrically \citep{Avin2005}.

\citet{Tchetgen2014} showed that identification can be retained even under treatment-dependent confounding by further assumptions such as monotonicity of the treatment effect on a binary treatment-dependent confounder.
Monotonicity here means that the treatment may have only a positive (or negative) effect on the treatment-dependent confounder.
Other assumptions retaining point identification include independence of counterfactuals of the treatment-dependent confounder or absence of an additive interaction between the mediator and the treatment-dependent confounder \citep{Robins2010, Tchetgen2014}.
Without any additional assumptions, the mediational effects under an observed discrete treatment-dependent confounder are still partially identifiable.
This means that lower and upper bounds for the estimates can be obtained \citep{Miles2017}.

Further methodological issues arise when study participants are followed over time and the mediator is a longitudinal process.
While causal mediation analysis has been extended to scenarios with longitudinal mediators \citep{Lin2017,VanderWeele2017,Zheng2017}, most literature treats the mediator as a vector-valued entity containing successive measurements of the mediating variable.
Such an approach results in high dimensionality if the number of repeated measurements per individual is large and also poses challenges in handling uneven measurement intervals and missing values. 
Some authors have instead applied functional regression to represent the mediator trajectory as a function \citep{Lindquist2012, Coffman2023, Zeng2023}. 
Treating the mediator as a functional entity avoids the problem of dimensionality and enables flexible use of its full history at the time the response is evaluated. 
The repeated measurements in the longitudinal setting can also be used to extract information about individual-level latent properties.
For example, \citet{Zheng2021} relaxed the assumption of no unmeasured mediator--outcome confounding under a longitudinal mediator and a time-to-event outcome by employing a joint modelling framework to estimate and control a common random effect reflecting an unobserved confounder between the two.

Time-to-event outcomes pose additional challenges in the causal inference framework \citep{Didelez2019}. 
A particular issue arises when causal estimands are defined by measures that condition on prior survival, such as the hazard function.
If there exist latent variables affecting survival, conditioning opens a backdoor path from the treatment to future survival through the latent variables.
This issue can be addressed by defining the response as an unconditional measure, such as the restricted mean survival time (RMST), i.e. the mean event-free time within a preset time period \citep{Zeng2023,Royston2013}.
If a significant portion of individuals do not experience the event during the study, RMST has the additional benefit over the mean survival time of allowing the time period to be chosen so that it remains robust to misspecification of the unobserved tail of the event-time distribution.
In addition, if it is considered possible to have a zero risk for the event, the true mean survival time would be infinite, whereas RMST would remain constrained to the chosen clinically relevant time period. 

In this study, we address identification of mediational causal effects in interventional studies with treatment-dependent confounding.
We extend the previously presented monotonicity assumption \citep{Tchetgen2014} to a trichotomous treatment-dependent confounder and show that this results in expressions identifiable up to a stratum-specific sensitivity parameter.
Our approach is similar to the partial identification in \citet{Miles2017} but imposes restrictions on the unobserved joint distribution of the counterfactuals of the treatment-dependent confounder, leading to a necessary condition for their marginals.
As the marginals can be estimated from observed data, the feasibility of the monotonicity assumption can be empirically assessed.
As an application, we consider the effect of an intensive lifestyle intervention on T2D incidence among a high-risk population, based on the Finnish Diabetes Prevention Study (DPS) \citep{Lindstrom2003}. 
The aim of the empirical analysis is to quantify the extent to which the effect of the intervention on T2D-free time is mediated through it inducing weight loss.
We apply functional regression analysis to represent the body mass index (BMI) trajectory as a functional entity and use a joint modelling framework to control potential latent confounding between the BMI trajectory and T2D incidence.

The paper is structured is as follows.
Section \ref{section_motivation}, presents the empirical problem motivating this study.
Section \ref{section_methods}, defines the targeted causal estimands and provides conditions for their identification and the methods we propose to use for their estimation.
Sections \ref{section_application} and \ref{section_discussion}, present the results of the empirical application and discuss the results along with further considerations of the used methodology and its possible limitations.

\section{Data sources and motivation}\label{section_motivation}

The aim of the Finnish Diabetes Prevention Study (DPS) is to assess the effectiveness of an intensive lifestyle intervention in preventing and delaying T2D onset in a high-risk population \citep{Eriksson1999,Tuomilehto2001,Lindstrom2003}. 
The study cohort was enrolled between 1993 and 1998 and originally consisted of 522 individuals.
The eligibility criteria required being overweight ($\hbox{BMI}>25$), aged 40--64 years and having impaired glucose tolerance at the screening visit.
The study participants were randomly allocated to intervention and control groups.
The active intervention lasted a maximum of six years (median 4 years), involving frequent personalised nutritional counselling and encouragement for physical activity, primarily through face-to-face sessions. 
The active intervention ended in 2001.
The control group was given routine, non-personalised healthy lifestyle advice during the study visits. 
Post-intervention follow-up visits continued until 2013, with a median of ten clinical study visits per person.
The participants were tested for T2D in the clinical study visits using the World Health Organization's 1985 criteria for a 2-hour oral glucose tolerance test (OGTT) \citep{WHO}, with a diagnosis requiring two OGTTs above the threshold.
In addition, the Finnish Registers for Drug Reimbursements and Drug Purchases were searched for T2D-related drug purchases, extending the time-to-event follow up until the end of 2018. 

A previous analysis of the DPS data showed a 40\% lower hazard for T2D and greater weight loss in the intervention group during the first 13 years of follow-up, and also showed greater improvements in their lifestyle compared with the control group, particularly in dietary intakes \citep{Lindstrom2013}.
Moreover, previous analyses have found associations between the lifestyle intervention, physical activity, nutritional components and diabetes incidence in the DPS cohort \citep{Laaksonen2005,Lindstrom2006}.

Since obesity is one of the most prominent risk factors for T2D, it is of interest to assess the extent to which the success of the lifestyle intervention can be attributed to its ability to reduce body weight in people with overweight or obesity.
Previous literature on lifestyle intervention studies has suggested a so-called legacy effect, wherein the intervention's impact on T2D incidence persist long after the intervention ends and the obtained group differences in risk factors have diminished.
As summarised by \citet{Wilding2014}, such results have been reported in the major T2D prevention trials, including the DPS \citep{Lindstrom2013}, the study conducted in China \citep{DaQinq} and the Diabetes Prevention Program in the USA \citep{dpp}.
These findings motivated us to consider the change in the BMI during the early phase of the intervention as the effective mediator.
Moreover, we apply a three-year time window because the majority of differences in BMI between the two groups occurred during this period.
We will use the restricted survival time as the outcome measure, choosing the maximum time as 15 years, reflecting a clinically relevant time period.
The outcome is thus interpreted as the number of healthy (i.e. T2D-free) years during the first 15 years after intervention onset.

In addition to weight reduction, the DPS intervention aimed at moderate physical activity and healthy nutritional composition measured by intakes of total fats, saturated fats and fibre \citep{Eriksson1999}.
These lifestyle factors can be assumed to influence the study participants' BMI trajectories and also T2D incidence through mechanisms other than weight loss, thus rendering them potential treatment-dependent confounders.
We created a summary variable to represent individuals' lifestyle choices influenced by the intervention, combining total physical activity and the dietary intake components.
The amount of total physical activity was measured by self-reports at every study visit and the components of dietary intake by three-day food diaries prior to the study visits for the first three (in addition to the baseline) study visits.
All variables from each study visit were standardised with respect to their baseline means and standard deviations and the lifestyle score was computed as the mean over the standardised variables over the three post-baseline study visits.
The lifestyle score was then categorised to three levels based on the observed tertiles of the score at baseline.

Causal mediation analysis requires controlling for any factors confounding the relationships between the treatment, mediator and outcome.
We considered age, sex, smoking status and the baseline lifestyle score as potential confounding baseline variables.
Age at baseline was categorised as $<45$, 45 to 59 and $\geq 60$ years, while the smoking status was categorised as `never', `former' or `current'.

\section{Methods}\label{section_methods}

In this section we present the proposed methodological framework.
Sections \ref{section_causal_estimands} and \ref{section_causal_model} outline the assumed causal model and the estimation targets.
In Section \ref{section_identification} we give assumptions sufficient to identify the causal estimands from empirical data and the resulting expressions for the direct and indirect effects.
In Sections \ref{section_parametric_models} and \ref{section_estimation} we define the parametric models and describe the strategy for their estimation.

\subsection{Causal estimands}\label{section_causal_estimands}
Let $A \in \{a^*, a\}$ denote the treatment group ($a$ for intervention, $a^*$ for control), $\widetilde{T}$ time since baseline (study onset) to T2D diagnosis, $T =\min\{\widetilde{T},t_{ \max }\}$ the restricted time without a T2D diagnosis for a prespecified $t_{\max}$ ($=15$ years), and $M(\cdot)$ a function of time that describes the true trajectory of BMI as a continuous mediator if remaining alive and T2D-free. 
We use subscripts to denote quantities under potential, possibly counterfactual scenarios: $T_a$ refers to the restricted survival time when the treatment is set to $a$, whereas $T_{a,M_{a^*}(\cdot)}$ is the corresponding time when the treatment is set to $a$, but the mediator follows the trajectory it would take under intervention $a^*$. 

We use the restricted mean survival time (RMST), $\tau^{t_{\max}} = \mathbb{E}(T^{(i)}) =\mathbb{E}\big( \int_0^{t_{\max}} P(\widetilde{T}^{(i)}>v)\dd{v}\big)$, as the response measure \citep{Royston2013}, and define the average natural direct and indirect effects of the intervention by contrasting treatment $a$ against $a^*$ as
\begin{equation}\label{eq_causal_effects}
\begin{aligned}
DE &= \mathbb{E}\Big[ T^{(i)}_{a,M_{a^*}(\cdot)} -  T^{(i)}_{a^*,M_{a^*}(\cdot)}  \Big] = \tau^{t_{\max}}_{a,M_{a^*}(\cdot)} - \tau^{t_{\max}}_{a^*,M_{a^*}(\cdot)}, \\
\\
IE &= \mathbb{E}\Big[ T^{(i)}_{a,M_{a}(\cdot)} -  T^{(i)}_{a,M_{a^*}(\cdot)}  \Big]  = \tau^{t_{\max}}_{a,M_{a}(\cdot)} - \tau^{t_{\max}}_{a,M_{a^*}(\cdot)}.
\end{aligned}
\end{equation}
The direct effect ($DE$) describes the change in RMST if the treatment was set to $a$ instead of $a^*$ but the mediator follows the trajectory it would have under treatment $a^*$. 
The indirect effect ($IE$) is the change in RMST if the mediator follows the trajectory it would have under treatment $a$ instead of treatment $a^*$ while the treatment is fixed to $a$ (see e.g. \citealp{VanderWeele2015}).
The direct and indirect effects sum up to the average total effect, $TE = \tau^{t_{\max}}_{a,M_a(\cdot)} - \tau^{t_{\max}}_{a^*,M_{a^*}(\cdot)}$.
Of note, the RMST corresponds to the area under the survival function between times zero and $t_{\max}$, implying that a simple nonparametric estimator of the total effect can be constructed using the areas under the Kaplan-Meier curves for the two treatment groups.

If there is interaction between the treatment and the mediator, the interpretation of mediational effects depends on the choice of the levels at which the fixed variables are held in each case.
The definitions \eqref{eq_causal_effects} lead to interpreting the direct effect as the \emph{pure direct effect} and the indirect effect as the \emph{total indirect effect}, as discussed by \citet{VanderWeele2013, VanderWeele2014}.
For ease of notation, we hereafter denote the mediator trajectory and its realisation often simply as $M$ and $m$, respectively.

\subsection{Causal model}\label{section_causal_model}
\begin{figure}[h!]
	\includegraphics[width=.5\linewidth]{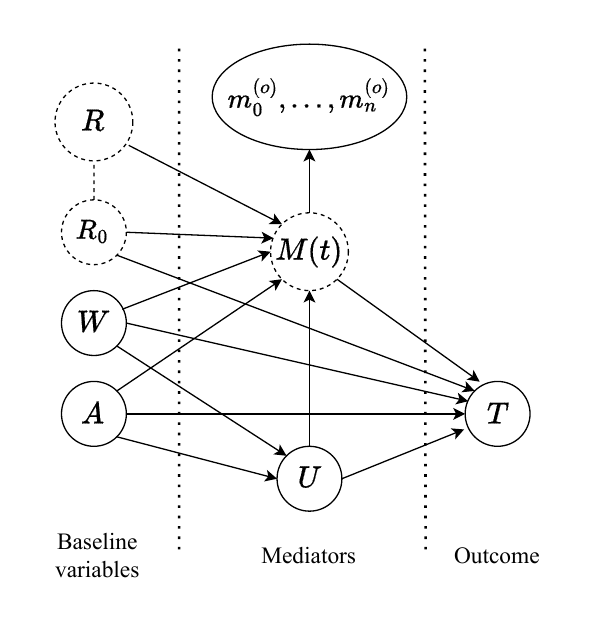}
	\centering
	\caption{Directed acyclic graph describing the assumed causal mechanism within an individual. The effect of treatment $A$ on the restricted survival time $T$ is assumed to be mediated through the underlying mediator trajectory $M(t)$ of which $n$ repeated measurements $m_1^{(o)},\dots,m_n^{(o)}$ are observed over time. The random effect $R_0$ is shared between the mediator $M(t)$ and the time-to-event outcome $T$, while the random effects $\boldsymbol{R}$ affect only the mediator. The dashed line between $R_0$ and $\boldsymbol{R}$ implies that they are correlated. $U$ is the treatment-dependent confounder and $\boldsymbol{W}$ consists of the baseline confounders. The dashed nodes refer to latent (unobserved) variables.}
	\label{fig_dag}
\end{figure}

The causal model is represented graphically by the \emph{directed acyclic graph} (DAG) shown in Figure \ref{fig_dag}, describing the assumed causal structure of the relevant variables involved.
The main interest lies in the interplay between the treatment $A$ (lifestyle intervention), the mediator trajectory $M$ (BMI trajectory) and the restricted survival time $T$ as the outcome. 
The paths from the treatment to the survival time bypassing the mediator represent the direct effect of the treatment on the outcome, while the indirect effect follows the paths traversing from the treatment to the survival time through the mediator.
The treatment-dependent confounder $U$ represents lifestyle behaviour that is influenced by the intervention but is not of primary interest.
The variable set $W$ consists of the confounding baseline variables that are considered to affect the mediator trajectory, the restricted survival time and the lifestyle behaviour.

We assume that there are individual-specific random effects ($R_0$,$\boldsymbol{R}$) that affect the outcome through the underlying mediator trajectory.
In addition, $R_0$ is allowed to have a separate, direct effect, constituting frailty for the time-to-event outcome, and furthermore, a confounder for the causal mechanism.
The random effects can be seen as some properties of the individuals which could in principle be measured and adjusted for and, although unobserved in reality, they can be learned through their influence on the repeated measurements of the mediator.
A similar approach has previously been adopted to estimate the mediational effects of certain drug treatments on overall survival mediated by CD4 cell count \citep{Zheng2021}.

We interpret the DAG in the framework of nonparametric structural equations models (NPSEM) \citep{Pearl2001}.
The directed arrows imply a causal ordering between the variables, and an absence of an arrow between two variables implies no direct causal relationship between them.
Importantly, the absence of any bidirectional arrows implies an assumption that any randomness affecting one variable in the graph is independent of the randomness affecting any other variable.

\subsection{Identification of the causal effects}\label{section_identification}
Under the NPSEM framework, the following conditional independence assumptions are implied by the causal DAG of Figure \ref{fig_dag}:
\begin{enumerate}
\item $T_{a,u,m} \ind \{A,U,M\} \text{  } | \boldsymbol{W},R_0$
\item $M_{a,u} \ind \{A,U\} \text{  } | \boldsymbol{W}$
\item $U_a \ind A \text{  } | \boldsymbol{W}$
\item $M_{a,u} \ind \{U_a,U_{a*}\} \text{  } | \boldsymbol{W}$
\item $T_{a,u,m} \ind \{U_a, U_{a^*},M_{a^*,u'}\} \text{  } | \boldsymbol{W},R_0$.
\end{enumerate}
The first three assumptions translate to no unmeasured confounding between $A$, $U$, $M$ and $T$.
The assumptions 4 and 5 are \emph{cross-world independencies}, essentially requiring that there exist no further confounders between $U$, $M$ and $T$ which are affected by the treatment.
In addition, we take the missing data mechanism to be missing at random, i.e. the right censorings of the survival process and drop-outs in the longitudinal follow-ups are considered non-informative.
Under these assumptions, the direct and indirect effects \eqref{eq_causal_effects} can be identified up to a term containing the joint distribution of $(U_a, U_{a^*})$.
We will show below that in the case of a trichotomous treatment-dependent confounder $U$ and assuming the monotonicity of the treatment effect on $U$ will further allow the identification of the estimands up to a stratum-specific sensitivity parameter.

Let $m^\dagger$ and $u^\dagger$ denote some reference trajectory of the mediator and a reference value of $U$, respectively.
Following \citet{Tchetgen2014}, the expected outcome, given $A$, $M$, $U$, $\boldsymbol{W}$ and $R_0$, can be decomposed into four terms representing the main effects of $M$ and $U$, the interaction between $M$ and $U$, and a reference level:
$$
\begin{aligned}
\beta_{m}(a,m,\boldsymbol{w},r_0) &= \mathbb{E}(T|a,m,u^\dagger,\boldsymbol{w},r_0) -  \mathbb{E}(T|a,m^\dagger,u^\dagger,\boldsymbol{w},r_0), \\
\beta_{u}(a,u,\boldsymbol{w}) &=  \mathbb{E}(T|a,m^\dagger,u,\boldsymbol{w}) - \mathbb{E}(T|a,m^\dagger,u^\dagger,\boldsymbol{w}), \\
\beta_{m,u}(a,m,u,\boldsymbol{w}) &= \mathbb{E}(T|a,m,u,\boldsymbol{w}) - \mathbb{E}(T|a,m^\dagger,u,\boldsymbol{w}), \\
& \quad -  \mathbb{E}(T|a,m,u^\dagger,\boldsymbol{w}) + \mathbb{E}(T|a,m^\dagger,u^\dagger,\boldsymbol{w}), \\
\bar{\beta}_{a,\boldsymbol{w}}(a,\boldsymbol{w}) &=  \mathbb{E}(T|a,m^\dagger,u^\dagger,\boldsymbol{w}).
\end{aligned}
$$
The `no additive interaction' assumption of Tchetgen Tchetgen and VanderWeele assumes that the terms $\beta_{m,u}(\cdot)$ are zero for all $a$, $m$, $u$ and $\boldsymbol{w}$. 
However, here we will retain these additive interaction terms so that the resulting expressions for the direct and indirect effects are (for details, see the online supplementary material)
\begin{equation}\label{eq_de_ie}
\begin{aligned}
DE &= DE^{(r)} - \Delta_{DE} + \delta, \\
IE &= IE^{(r)} + \Delta_{IE} - \delta,
\end{aligned}
\end{equation}
where
\begin{equation*}
\begin{aligned}
DE^{(r)} &=  \int_{m,\boldsymbol{w},r_0}\big[ \beta_m(a,m,\boldsymbol{w},r_0) - \beta_m(a^*, m,\boldsymbol{w},r_0) \big] P_M(m|a^*,\boldsymbol{w},r_0) \times \\
& \qquad \qquad P_{\boldsymbol{W}}(\boldsymbol{w})P_{R_0}(r_0) \dd{m}\dd{\boldsymbol{w}}\dd{r_0}\text{ } + \\
& \quad \int_{u,\boldsymbol{w}}\big[ \beta_u(a,u,\boldsymbol{w})P_U(u|a,\boldsymbol{w}) - \beta_u(a^*,u,\boldsymbol{w})P_U(u|a^*,\boldsymbol{w}) \big]P_{\boldsymbol{W}}(\boldsymbol{w}) \dd{u}\dd{\boldsymbol{w}} \text{ } + \\
& \quad  \int_{\boldsymbol{w}} \big[ \bar \beta_{a,\boldsymbol{w}}(a,\boldsymbol{w}) - \bar \beta_{a,\boldsymbol{w}}(a^*,\boldsymbol{w}) \big]P_{\boldsymbol{W}}(\boldsymbol{w})\dd{w}, \\
IE^{(r)} &= \int_{m,\boldsymbol{w},r_0}  \beta_m(a,m,\boldsymbol{w},r_0) \big[ P_M(m|a,\boldsymbol{w},r_0)-P_M(m|a^*,\boldsymbol{w},r_0) \big] \times \\
& \qquad \qquad P_{\boldsymbol{W}}(\boldsymbol{w}) P_{R_0}(r_0)\dd{m} \dd{\boldsymbol{w}}\dd{r_0}, \\
\Delta_{DE} &= \int_{m,u',\boldsymbol{w},r_0}\beta_{m,u}(a^*,m,u',\boldsymbol{w},r_0)P_M(m|a^*,u',\boldsymbol{w},r_0)P_U(u'|a^*,\boldsymbol{w}) \times \\
& \qquad \qquad P_{\boldsymbol{W}}(\boldsymbol{w})\dd{m}\dd{u'}\dd{r_0}, \\ 
\Delta_{IE} &=  \int_{m,u,\boldsymbol{w},r_0}\beta_{m,u}(a,m,u,\boldsymbol{w},r_0)P_M(m|a,u,\boldsymbol{w},r_0)P_U(u|a,\boldsymbol{w}) \times \\ 
& \qquad \qquad P_{\boldsymbol{W}}(\boldsymbol{w})P_{R_0}(r_0) \dd{m}\dd{u}\dd{\boldsymbol{w}}\dd{r_0},\\
\delta &= \int_{m,u,u',\boldsymbol{w},r_0}\beta_{m,u}(a,m,u,\boldsymbol{w},r_0)P_M(m|a^*,u',\boldsymbol{w},r_0)P(U_a=u,U_{a^*}=u'|\boldsymbol{w}) \times \\
&\qquad \qquad P_{\boldsymbol{W}}(\boldsymbol{w})P_{R_0}(r_0) \dd{m}\dd{u}\dd{u'}\dd{\boldsymbol{w}}\dd{r_0}.\\
\end{aligned}
\end{equation*}
The part of the total effect that is transmitted through the additive interaction of $M$ and $U$ on $T$ is given by $\Delta_{IE}-\Delta_{DE}$, which is identifiable.
The term $\delta$ controls how this effect is divided into the direct and indirect effects but is not itself identifiable as its expression relies on the joint probability of counterfactual levels of $U$.
This also means that the direct and indirect effects are not identifiable without further assumptions.

Tchetgen Tchetgen and VanderWeele showed that assuming no additive interaction of $M$ and $U$ on $T$, or, in the case of a binary $U$, monotonicity of the effect of $A$ on $U$, would be enough to identify $\delta$ and, therefore, also the direct and indirect effects from empirical data \citep{Tchetgen2014}.
The `no additive interaction' assumption would imply $\Delta_{DE}=\Delta_{IE} = \delta = 0$, and subsequently $DE=DE^{(r)}$ and $IE = IE^{(r)}$.
The monotonicity assumption of the effect of $A$ on $U$ means that an individual cannot have a worse value of $U$ under treatment than they would have had under no treatment.
If $U$ is binary, the joint probability of $U_a$ and $U_{a^*}$ becomes fully determined by the marginal probabilities, which can be estimated from the observed data.

We now extend the monotonicity assumption to a trichotomous $U$ and show that, in each of the strata defined by the baseline covariates, the joint probability is identified up to a sensitivity parameter. 
For convenience, we omit denoting the stratum in what follows.
Suppose that $U$ is trichotomous, taking values $U\in\{0,1,2\}$, and the treatment effect on $U$ is monotonic.
The stratum-specific joint probability of $(U_a, U_{a^*})$ can then be represented in terms of six probability parameters:

\begin{center}
\begin{tabular}{c c|c c c | c}
& & & $U_a$ & & \\
& & 0 & 1 & 2 & \\
\hline
&2 & 0 & 0 & $p_{22}$ & $\Phi_6$ \\
$U_{a^*}$ & 1& 0 & $p_{11}$ & $p_{12}$ & $\Phi_5$ \\
& 0 & $p_{00}$ & $p_{01}$ & $p_{02}$ & $\Phi_4$ \\
\hline
& & $\Phi_1$ & $\Phi_2$ & $\Phi_3$ & 
\end{tabular}
\end{center}
where the marginal probabilities $\Phi$ can be estimated (i.e. are identifiable) from the observations.
We obtain directly $p_{00} = \Phi_1 = P(U=0|a)$ and $p_{22} = \Phi_6 = P(U=2|a^*)$.
Conditionally on the marginals, there is thus only one degree of freedom for the remaining four parameters.
Considering $p_{11}$ as the free parameter, it is constrained by the marginals to the interval
\begin{equation}\label{eq_p11}
p_{min} = \max\{0,1-\Phi_3 - \Phi_4 \} \leq p_{11} \leq \min\{\Phi_2,\Phi_5\}=p_{max}.
\end{equation}
The admissible values for $p_{11}$ can now be expressed in terms of a sensitivity parameter $\rho = (p_{11}-p_{min})/(p_{max}-p_{min}) \in[0,1]$, describing where $p_{11}$ lies within its supported interval.
For a fixed $\rho$, the term $\delta$ is identified and its range of possible values can be calculated by varying the sensitivity parameter $\rho$ within the $[0,1]$ interval.
The lower and upper bounds for the direct and indirect effects can be found by minimising and maximising $\delta$ within each stratum.
Since each $\delta$ is linear in the probabilities $p_{ij}$, its minimum and maximum values for a given stratum are found at the opposite ends of the $\rho$ interval.
Furthermore, \eqref{eq_p11} implies that $1-\Phi_3-\Phi_4 \leq \min\{\Phi_2,\Phi_5\}$ is a necessary (but not sufficient) condition for the monotonicity assumption to hold.
The marginal probabilities $\Phi$ can be estimated from data and thus allow one to assess based on empirical data whether the monotonicity assumption can be considered viable.
Generalisation to the case with a treatment-dependent confounder with arbitrary number of levels is given in the online supplementary material.

If $p_{11}=1-\Phi_3 - \Phi_4$, then $p_{02}=0$ and we obtain a special case, here referred to as \emph{step monotonicity}.
This assumption means that the treatment can either have no effect on $U$, or elevate $U$ to one level higher.
Under step monotonicity, the joint probability for $U_a$ and $U_{a^*}$ is identifiable by the marginals without the need of the sensitivity parameter, provided that the marginals are consistent with the step monotonicity.
Of note, under step monotonicity the identifiablity holds for any ordinal variable $U$ with any number of possible values (for details, see the online supplementary material).

\subsection{Parametric models}\label{section_parametric_models}
Under the identifying assumptions of Section \ref{section_identification}, the direct and indirect effects can be expressed in terms of the observed data.
Although the effects and their corresponding empirical expressions were derived nonparametrically, the components forming the empirical expressions were estimated parametrically.
In this section we describe the parametric models we used to estimate the terms in the expressions \eqref{eq_de_ie} for the direct and indirect effects.

\subsubsection*{Mediator trajectory}
We assumed a linear mixed model for the mediator trajectory. 
The underlying true mediator was assumed to be a smooth trajectory $M_i(t)$ from which observations are made with stochastic deviations.
In particular, the observed longitudinal measurements $m_{ij}^{(o)}$ for individual $i$ were assumed to arise from the model
\begin{equation}\label{eq_longmodel}
\begin{aligned}
m_{ij}^{(o)} &= M_i(t_{ij}) + \epsilon_{ij}, \quad j  = 1,\dots,n_i, \\
M_i(t) &= (\beta_0+R_{i0}) + \beta_1' \boldsymbol{X}_i +  \beta_2 ' \boldsymbol{W}_i + \sum_{k=1}^4 (\alpha_k+ \psi_k' \boldsymbol{X}_i) B_k(t)+ \sum_{k=1}^{3}R_{ik}B^{r}_{k}(t),
\end{aligned}
\end{equation}
where $\boldsymbol{X}_i = (A_i,I(U_i=1),I(U_i=2),A_i I(U_i=1), A_i I(U_i=2))'$ and $\boldsymbol{W}_i$ is a vector containing the baseline covariates (age, sex, smoking status and lifestyle score).
The terms $B_k$ are population-level basis functions for natural cubic splines with outer knots placed at the baseline and 10 years and three inner knots placed at 1, 3 and 5 years since study onset.
Similarly, $B^r_k$ are basis functions corresponding to individual-level random effects, with two inner knots placed at years 1 and 5.
The $\epsilon_{ij}$ are normally distributed mutually independent error terms with variance $\sigma^2$ and the random effects $(R_{i0},R_{i1},R_{i2},R_{i3})'$ have a multinormal distribution with mean zero and a full covariance matrix $\Sigma$.
Note that the model includes time-independent effects of the treatment and treatment-dependent confounder on the outcome.
This entails the assumptions that the treatment-dependent confounder is affected by the treatment without delay, and it in turn affects the mediator trajectory without delay.

\subsubsection*{Time-to-event outcome}
We assumed a parametric proportional hazards model for the time-to-event outcome with separate piecewise-constant baseline hazards for the two treatment groups.
Denoting the baseline hazard functions as $h_{00}(t)$ and $h_{01}(t)$, the hazard for individual $i$ was modelled as
\begin{equation}\label{eq_survmodel}
h_i(t) = h_{00}(t)^{(1-A_i)}h_{01}(t)^{A_i}\exp \big\{ \boldsymbol{\gamma}_1'\boldsymbol{U}_i + \boldsymbol{\gamma}_2'\boldsymbol{U}_iA_i + \boldsymbol{\gamma}_3' \boldsymbol{W}_i +  g(M_i(\cdot), t)\boldsymbol{G}_i'\boldsymbol{\zeta} + \xi R_{i0}  \big\},
\end{equation}
where $g(\cdot)$ is a function of the full mediator trajectory evaluated at $t$ and determines the parametric form of the dependency between the mediator process and the hazard function.
$\boldsymbol{G}_i$ is the vector $(1,\boldsymbol{U}_i',A_i)'$, meaning that parameter $\zeta_1$ is interpreted as the main effect of the mediator, and the rest as interaction effects between the mediator and the levels of the treatment-dependent confounder and the treatment.
The individual-level random intercept $R_{i0}$ from the longitudinal submodel enters the hazard function as a frailty, with the parameter $\xi$ describing the strength of its effect on the hazard, thus controlling for latent confounding of the mediator--outcome-relationship.

The choice of the functional form of $g(\cdot)$ should depend on the biological mechanism by which the mediator is assumed to affect the outcome.
Here, we choose a three-year legacy parameterisation, i.e. $g(M_i(\cdot),t) = \int_0^3 \big[M_i(v)-M_i(0)\big]\dd{v}$ for all $t\geq0$, implying that the hazard at any time is affected by the cumulative change of the latent trajectory over the first three years since the baseline.
We return to discuss the interpretation of this choice in the Discussion.
For comparison, we also consider a current change parameterisation where the hazard at time $t$ is affected by the change by time $t$ of the level of the latent trajectory since the baseline, i.e. $g(M_i(\cdot),t) = M_i(t) - M_i(0)$.

The RMST for an individual $i$ can be derived from the hazard function as
\begin{equation}\label{eq_rmst}
\tau^{t_{\max}}_i  = \int_0^{t_{\max}} P(\widetilde{T}^{(i)}>v)\dd{v} = \int_0^{t_{\max}} \exp\Big(-\int_0^{v} h_i(v')\dd{v'}\Big)\dd{v}.
\end{equation}
The parameters of the proportional hazards model should not themselves be interpreted as causal effects. 
Such an interpretation would require the implausible assumption that any source of between-individual heterogeneity would be accounted for in the model, for otherwise the parameters would suffer from selection bias due to unobserved heterogeneity \citep{Aalen2015}.

\subsubsection*{Treatment--dependent confounder}
We used multinomial logistic regression to model the dependence of the trichotomous treatment-dependent confounder $U$ on treatment $A$ and the baseline confounders $\boldsymbol{W}$. The log-ratios of probabilities for belonging to the categories $1$ or $2$ compared to the reference category $0$ were determined by
\begin{equation}\label{eq_Umodel}
\begin{aligned}
\log\Big( \frac{P(U=1|A,\boldsymbol{W})}{P(U=0|A,\boldsymbol{W})} \Big) &= \phi_0^{(1)} + \phi_1^{(1)}A + \phi_2^{(1)} \boldsymbol{W} =: \varphi^{(1)}(A,\boldsymbol{W}), \\
\log\Big( \frac{P(U=2|A,\boldsymbol{W})}{P(U=0|A,\boldsymbol{W})} \Big) &= \phi_0^{(2)} + \phi_1^{(2)}A + \phi_2^{(2)} \boldsymbol{W} =: \varphi^{(2)}(A,\boldsymbol{W}).
\end{aligned}
\end{equation}
The marginal probabilities $\Phi_1,\dots,\Phi_6$ of the joint distribution of $(U_a,U_{a^*})$ in a given stratum $\boldsymbol{W}$ were obtained by applying the inverse-logit transform, for example, $\Phi_2 = \exp\{\varphi^{(1)}(a, \boldsymbol{W})\}/(1+\exp\{\varphi^{(1)}(a, \boldsymbol{W}) \}+\exp\{\varphi^{(2)}(a, \boldsymbol{W})\})$.

The above model includes only the main effects of each predictor and could thus be considered relatively inflexible.
As the research question is concerned with the mediating mechanism of the treatment, we also considered models with interaction terms between the treatment and the baseline covariates to allow the treatment effect on the treatment-dependent confounder to differ among the baseline covariate strata.
Model comparison was carried out to determine whether any of the more flexible models should be favoured against model \eqref{eq_Umodel}.

\subsubsection*{Parametric causal effects}
Since the underlying BMI trajectory as the mediator is stripped of the stochastic error terms, its distribution conditionally on the covariates and treatment is determined by the distribution of the random effects. 
Integrating an arbitrary functional $q(m)$ over the possible realisations of $m$, given $A$, $U$ and $\boldsymbol{W}$, thus reduces to integrating over the joint distribution of the random effects.

With the definitions of mediational causal effects in \eqref{eq_causal_effects}, the assumed parametric models imply the following formulae for the terms determining the direct and indirect effects:
\begin{equation*}
\begin{aligned}
DE^{(r)} &= \int_{\boldsymbol{r}} \sum_{\boldsymbol{w},u'} \big[ \beta_m(a,m(a^*,u',\boldsymbol{r},\boldsymbol{w}),r_0,\boldsymbol{w}) - \beta_m(a^*, m(a^*,u',\boldsymbol{r},\boldsymbol{w}),r_0,\boldsymbol{w}) \big]\times \\ & \hspace{.2\linewidth}  P_{\boldsymbol{R}}(\boldsymbol{r})P_U(u'|a^*,\boldsymbol{w})P_{\boldsymbol{W}}(\boldsymbol{w}) \dd{\boldsymbol{r}}\text{ } + \\
& \quad \sum_{u,\boldsymbol{w}}\big[ \beta_u(a,u,\boldsymbol{w})P_U(u|a,\boldsymbol{w}) - \beta_u(a^*,u,\boldsymbol{w})P_U(u|a^*,\boldsymbol{w}) \big]P_{\boldsymbol{W}}(\boldsymbol{w}) \text{ } + \\
& \quad  \int_{r_0}\sum_{\boldsymbol{w}} \big[ \bar \beta_{a,c}(a,\boldsymbol{w},r_0) - \bar \beta_{a,c}(a^*,\boldsymbol{w},r_0) \big]P_{R_0}(r_0)P_{\boldsymbol{W}}(\boldsymbol{w})\dd{r_0}, \\
\\
IE^{(r)} &=   \int_{\boldsymbol{r}} \sum_{\boldsymbol{w},u,u'} \Big[ \beta_m(a,m(a,u,\boldsymbol{r},\boldsymbol{w}),r_0,\boldsymbol{w})P_U(u|a,\boldsymbol{w}) - \\ 
& \hspace{.2\linewidth}\beta_m(a,m(a^*,u',\boldsymbol{r},\boldsymbol{w}),r_0,\boldsymbol{w})P_U(u'|a^*,\boldsymbol{w}) \Big] P_{\boldsymbol{R}}(\boldsymbol{r})P_{\boldsymbol{W}}(\boldsymbol{w})\dd{\boldsymbol{r}},\\ \\
\Delta_{DE} &= \int_{\boldsymbol{r}}\sum_{u',\boldsymbol{w}}\beta_{m,u}(a^*,m(a^*,u',\boldsymbol{r},\boldsymbol{w}),u',\boldsymbol{w})P_{\boldsymbol{R}}(\boldsymbol{r})P_U(u'|a^*,\boldsymbol{w})P_{\boldsymbol{W}}(\boldsymbol{w})\dd{\boldsymbol{r}}, \\ 
\Delta_{IE} &=  \int_{\boldsymbol{r}}\sum_{u,\boldsymbol{w}}\beta_{m,u}(a,m(a,u,\boldsymbol{r},\boldsymbol{w}),u,\boldsymbol{w})P_{\boldsymbol{R}}(\boldsymbol{r})P_U(u|a,c)P_{\boldsymbol{W}}(\boldsymbol{w})\dd{\boldsymbol{r}}, \\
\delta &= \int_{\boldsymbol{r}}\sum_{u,u',\boldsymbol{w}}\beta_{m,u}(a,m(a^*,u',\boldsymbol{r},\boldsymbol{w}),u,r_0,\boldsymbol{w})P_{\boldsymbol{R}}(\boldsymbol{r})P(U_a=u,U_{a^*}=u'|\boldsymbol{w}) \times \\ 
& \hspace{.2\linewidth} P_{\boldsymbol{W}}(\boldsymbol{w})\dd{\boldsymbol{r}}.
\end{aligned}
\end{equation*}
Fully parametric expressions can then be obtained by plugging in the assumed joint distribution for the random effects, the parametric forms of $P_U(u|a,\boldsymbol{w})$ as implied by model \eqref{eq_Umodel}, and $\beta(\cdot)$ as implied by \eqref{eq_rmst} and \eqref{eq_survmodel}.

\subsection{Estimation}\label{section_estimation}
We employed a joint modelling framework to estimate the parametric models of the mediator trajectory and the time-to-event outcome.
Since these models rely on only the observed values of the treatment-dependent confounder, the model for the treatment-dependent confounder was estimated separately.
Pareto smoothed importance sampling leave-one-out cross-validation \citep{Vehtari2015} was used to compare the treatment-dependent confounder model against its more flexible variations.

A joint model for the longitudinal and time-to-event outcomes comprises specifying submodels for both outcomes and linking them via some association structure, thus allowing incorporating any information shared between the two outcomes \citep{Gould2015, Ibrahim2010, Hickey2018, Papageorgiou2019}.
The association structure was here induced by including a function of the longitudinal mediator trajectory into the linear predictor of the survival submodel and also assuming a random effect which is shared between the mediator trajectory and the time-to-event outcome enabling adjustment for a latent confounder.

We adopted a Bayesian framework to estimate all parametric models.
\sloppy Let $\boldsymbol{\theta}= (\boldsymbol{\beta}, \boldsymbol{\alpha}, \boldsymbol{\psi}, \sigma, \Sigma, \boldsymbol{\gamma}, \boldsymbol{\zeta}, \xi, h_{00},h_{01})$ denote all population-level model parameters in the joint model.
The joint distribution of observed longitudinal measurements $\boldsymbol{m}_i^{(o)}$, event time $T_i^{\text{exit}}$, event indicator $d_i$, and random effects $\boldsymbol{R}_i$ can be factorised as
\begin{equation*}
\begin{aligned}
p(\boldsymbol{m}_i^{(o)}, T_i^{\text{exit}},d_i,\boldsymbol{R}_i|A_i,\boldsymbol{W}_i ;\boldsymbol{\theta}) &= p(\boldsymbol{m}_i^{(o)}|\boldsymbol{R}_i,A_i,U_i,\boldsymbol{W}_i; \boldsymbol{\theta}) \times \\
& \quad p(T_i^{\text{exit}},d_i|\boldsymbol{R}_i,A_i,U_i,\boldsymbol{W}_i; \boldsymbol{\theta})p(\boldsymbol{R}_i;\boldsymbol{\theta}) \\
&=: p_i^m \times p_i^s \times p_i^r
\end{aligned}
\end{equation*}
The underlying mediator trajectory entering the survival submodel is determined by the random effects and the baseline covariates, making the survival observations conditionally independent of the actual measurements of the longitudinal process (see Figure \ref{fig_dag}).
The posterior distribution of the parameter vector $\boldsymbol{\theta}$ is
\begin{equation*}
\begin{aligned}
p(\boldsymbol{\theta} | \text{data}) &\propto \Big( \prod_i p_i^{m} \times p_i^{s} \times p_i^{r} \Big) \times p(\boldsymbol{\theta}), \text{ where}\\
p_i^{m} &\propto \sigma^{-n_i/2}\exp\{-\frac{1}{2} [\boldsymbol{m}_i^{(o)}-\mathbb{E}(\boldsymbol{m}_i^{(o)})]'[m_i^{(o)}-\mathbb{E}(\boldsymbol{m}_i^{(o)})]\sigma^{-2} \}, \\
p_i^{s} &= h_i(t_i^{\text{exit}})^{d_i}\exp\{ -\int_0^{t_i^{\text{exit}}} h_i(u) \dd{u} \},\\
p_i^{r} &\propto \det(\Sigma)^{-1/2} \exp\{-\frac{1}{2} \boldsymbol{R}_i'\Sigma^{-1}\boldsymbol{R}_i\},
\end{aligned}
\end{equation*}
and $p(\boldsymbol{\theta})$ is the prior distribution for $\boldsymbol{\theta}$.
The cumulative hazard in $p_i^s$ does not, in general, have a convenient analytical form, as it may involve a time-dependent functional of the mediator trajectory, and numerical integration is then needed for its computation.

For all of the regression parameters, we assumed relatively uninformative normal priors with mean zero and standard deviaton  $5$. 
As the longitudinal BMI measurements were centered at $25kg/m^2$ and scaled to one fifth of the original scale, effect sizes greater than $5$ would be implausible. 
Similarly, log hazard ratios of such magnitudes would be considered unrealistic. 
For the piecewise constant baseline hazards, we used $\text{Gamma}(.5,.5)$ priors for each piece. 
For the standard deviations of the residual terms in the longitudinal submodel and the random effects, half-Cauchy distributions were used with location parameter zero, and scale parameter $10$. 
In addition, a Lewandowski-Kurowicka-Joe prior was assigned to the Cholesky factor of the random effects correlation matrix.
The parameters $\psi$ in the multinomial model for the treatment-dependent confounder were given normal priors with mean $0$ and standard deviation $5$.
The models were run using four chains with 2000 burn-in iterations and 2000 sampling iterations each, resulting in 8000 Markov chain Monte Carlo (MCMC) draws from the posterior distribution of the model parameters and the convergence was assessed by the Gelman-Rubin $R$-statistic \citep{Gelman1992}, which was at most 1.01 for all parameters.
The implementation of the joint model was checked by simulating 100 datasets with parameters similar to those obtained with the observed data and checking that the model adequately recovered those parameters.
All models were fitted using the \emph{R} interface to the \emph{Stan} software \citep{Stan}, and post-processing was conducted using the \emph{R} software \citep{R}.

\section{Application}\label{section_application}
The DPS originally followed a cohort of 522 individuals, with 265 randomised to the intervention group and 257 to the control group.
After excluding individuals with missing values on the baseline covariates ($n=19$), the final sample consisted of 503 individuals with 254 people in the intervention group and 249 in the control group.
The median number of clinical visits per person was 11 (interquartile range 6--13) in the intervention group and 9 (IQR 4--13) in the control group.
The intervention group contributed a total of 3505 person years during which 166 T2D cases were observed with a restricted mean survival time of 11.2 years. The control group contributed a total of 2865 person years and 168 T2D cases with a restricted mean survival time of 9.5 years.

\begin{figure*}[!t]
	\includegraphics[width=\linewidth]{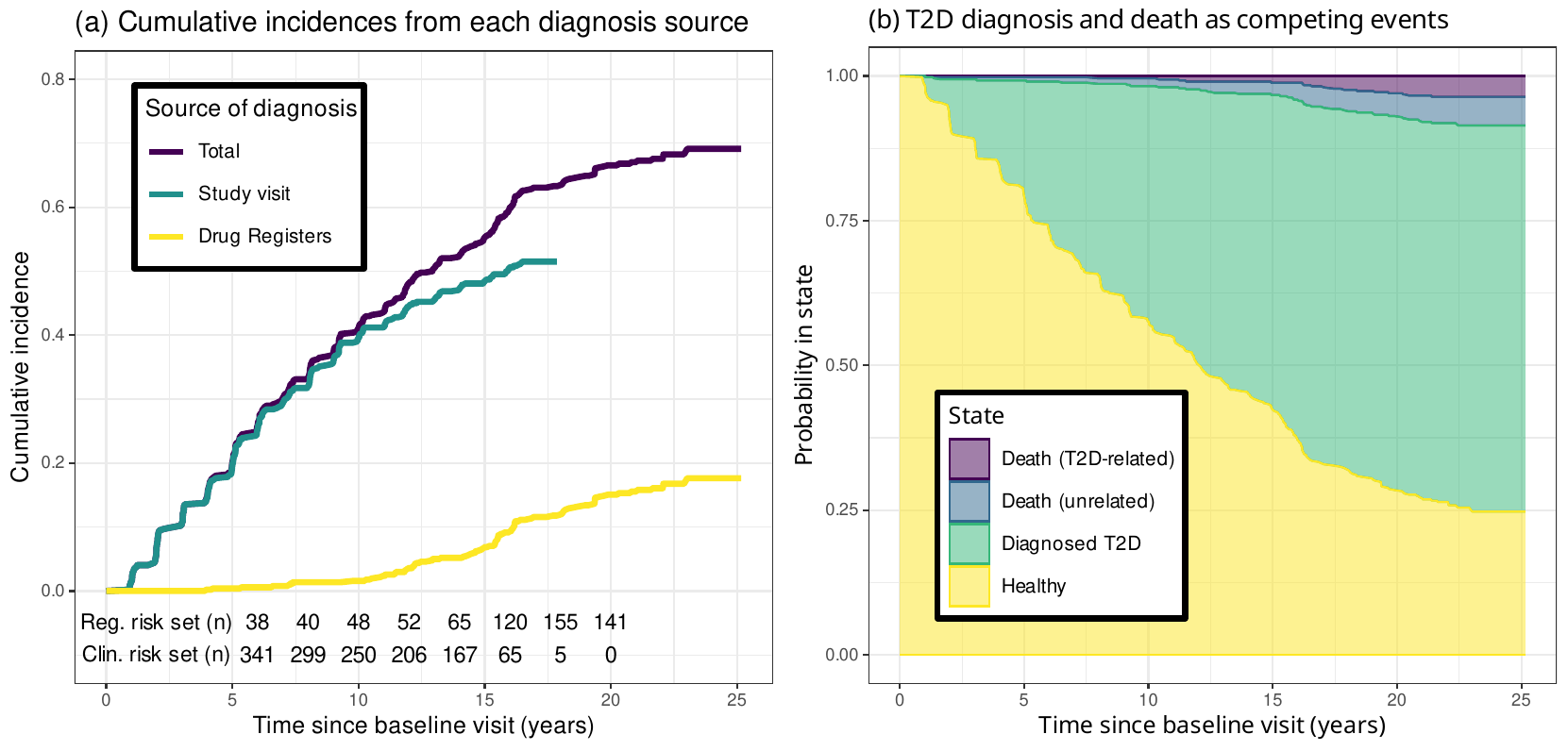}
	\centering
	\caption{ \textbf{(a)}: Cumulative incidence of T2D diagnoses due to both sources of diagnosis. Underneath the curves are shown the number of individuals being followed through each source. \textbf{(b)}: Probabilities of reaching the three competing (absorbing) states as functions of time. T2D-related deaths include cases where the cause of death is potentially associated with type 2 diabetes, namely cardiovascular and cerebrovascular complications.}
	\label{fig_endpoints}
\end{figure*}

The study endpoint was the diagnosis of T2D, either ascertained at any of the study visits or inferred from the register data.
Since the clinical study visits were considered the more reliable source, we used the first diagnosis made at the study visits as the primary endpoint and considered the register data only after their last study visit for each individual.
Figure \ref{fig_endpoints}($a$) shows the cumulative incidences of T2D from both sources.
The \emph{clinical risk set} refers to the number of individuals in the risk set having not yet made their last study visit, whereas the \emph{register risk set} is the number of individuals in the risk set being followed through the drug registers.
The cumulative incidence curves based on the study visits and the drug registers grow reasonably closely in proportion to the number of individuals, implying that any bias due to uneven sensitivity of T2D detection was unlikely.

For each study participant, the follow-up started at the baseline visit and terminated at the event of T2D diagnosis, death or end of follow-up at the end of 2018. 
The dates and causes of deaths were obtained from the Finnish Cause of Death Register. 
Figure \ref{fig_endpoints} (b) shows the cumulative risks for the competing events of T2D diagnosis, death with a potentially T2D-related cause, i.e. cardiovascular and cerebrovascular complications, and death from other causes. 
As the proportion of potentially T2D-related deaths appears negligible, treating all deaths as uninformative right censorings was deemed justified.
With the availability of the register data after the clinical follow-up, deaths were the only source of censoring in the data.

\subsection{Model fit}

\begin{figure}[h!]
	\includegraphics[width=\linewidth]{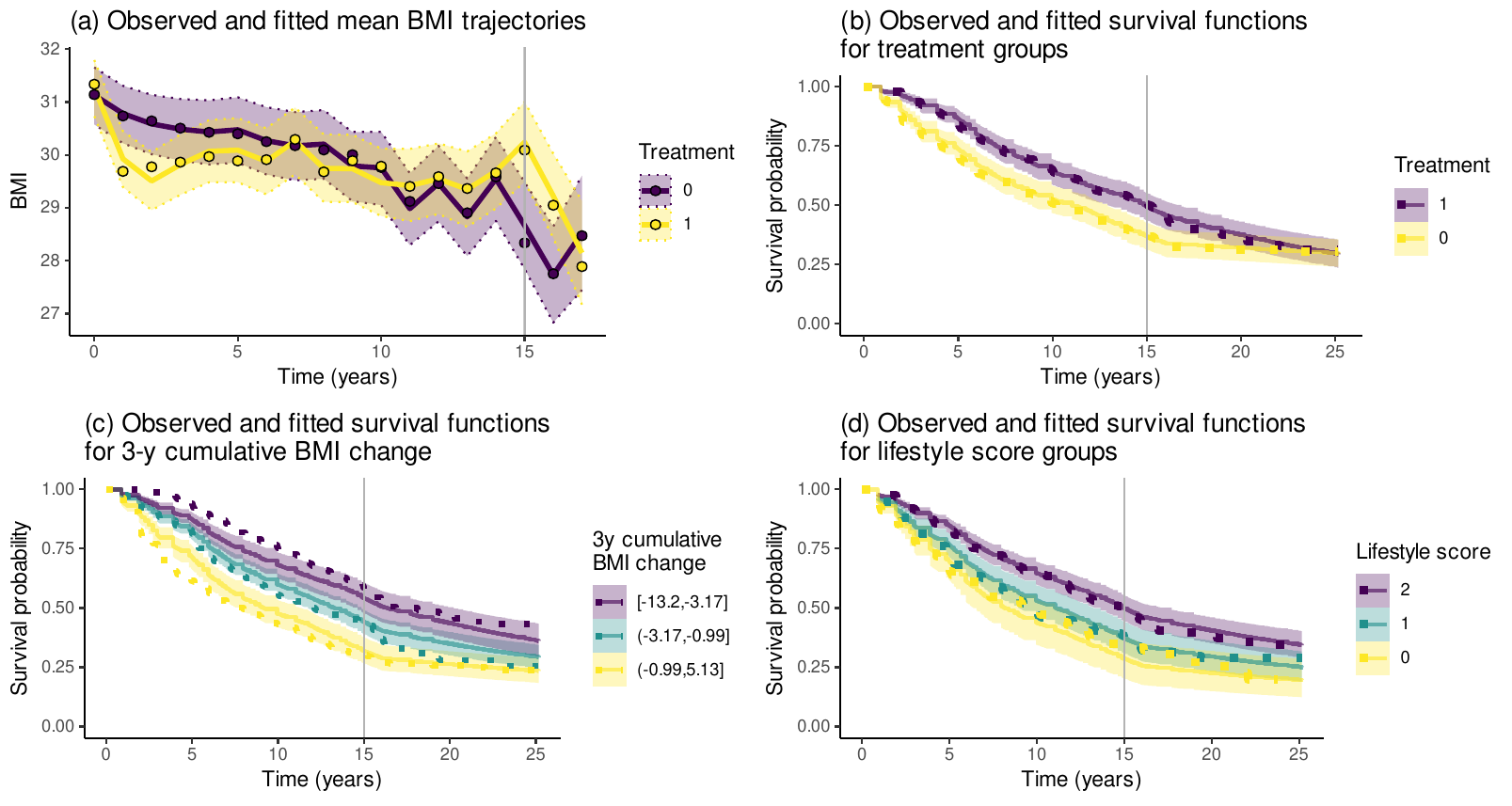}
	\centering
	\caption{ \textbf{(a)}: The observed (dots) and estimated (solid lines) mean trajectories of BMI in the intervention (1) and control (0) groups with the $95\%$ credible intervals.  \textbf{(b)}--\textbf{(d)}: Kaplan-Meier curves illustrating the effects of the treatment, early BMI change and lifestyle score on avoiding type 2 diabetes. The early BMI change is here categorised into the model-implied tertiles. The dotted lines are the Kaplan-Meier estimates, and the solid lines and the accompanying $95\%$ credible intervals are the estimated survival functions.}
	\label{fig_expl}
\end{figure}

Figure \ref{fig_expl}(a) shows the observed and estimated mean BMI trajectories for the two treatment groups.
Figures \ref{fig_expl}(b) to \ref{fig_expl}(d) display the Kaplan-Meier curves and estimated survival functions, separately illustrating the effects of the treatment, change in BMI over the first three years and lifestyle score on avoiding T2D.
In these plots, the cumulative change in BMI over the first three years was categorised into tertiles of the individual level point estimates obtained from the model.
The survival functions were then computed as the means of the estimated survival functions for each of the three groups.
These plots are based on the model using the three-year legacy parameterisation

The mean BMI trajectory in the intervention group (Figure \ref{fig_expl}(a)) exhibits a decrease in the early years since the start of the intervention, after which the mean trajectory gradually rebounds close to the trajectory of the control group.
The estimated curves replicate the observed average trajectories reasonably well, although the shape slightly differs in the intervention group over the first two years.
The Kaplan-Meier curves indicate a lower T2D risk for individuals in the intervention group (Figure \ref{fig_expl}(b)), in the higher lifestyle score groups (Figure \ref{fig_expl}(d)) or having decreased their BMI more over the first three years (Figure \ref{fig_expl}(c)).
The estimated survival functions agree closely with the Kaplan-Meier curves.
However, with the early BMI change the differences in the estimated survival functions are less pronounced than is evident from the Kaplan-Meier curves, which might indicate some lack of fit in the model with respect to the relationship between the early BMI change and T2D hazard.

\subsection{Model comparison and assessment of the monotonicity assumption}

In addition to model \eqref{eq_Umodel} of the treatment-dependent confounder, we considered more flexible models including interactions of the treatment with each baseline covariate all at the same time or each one separately.
The model comparison showed no discernible difference between the model's performances and so we chose the most parsimonious one, i.e., model \eqref{eq_Umodel} (see online supplement).

\begin{figure}[h!]
	\includegraphics[width=\linewidth]{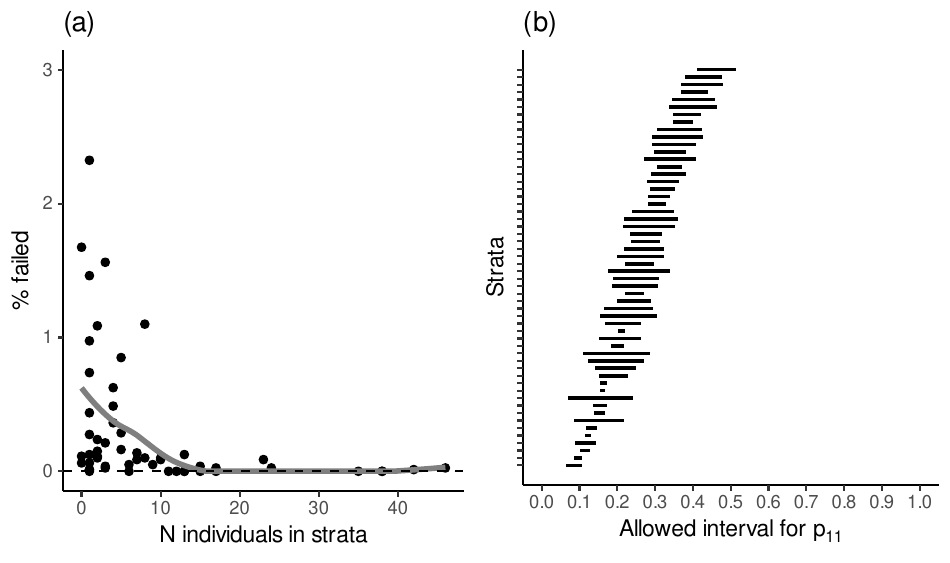}
	\centering
	\caption{$(a)$ Proportions of Markov chain Monte Carlo samples failing the monotonicity condition in the 54 strata defined by the covariates under the selected model. The largest proportions of failures occurred in strata with low numbers of individuals. (b) The average posterior boundaries within which the free parameter in the joint distribution of the two counterfactual treatment-dependent confounders ($p_{11}$) was constrained by the monotonicity assumption in each stratum.}
	\label{fig_sens}
\end{figure}

The monotonicity condition under the selected model was assessed from the empirical data by computing the estimated marginal probabilities for the two counterfactual treatment-dependent confounders ($U_a$ and $U_{a^*}$) for each of the 54 strata formed by the baseline covariates and checking whether the marginals satisfied the necessary condition for the monotonicity assumption.
Out of the 8000 MCMC samples, only 205 ($2.6\%$) contained one or more strata for which the marginals were not consistent with the monotonicity assumption.
Figure \ref{fig_sens}$(a)$ shows the proportion of these failures for each stratum and the number of individuals belonging to each in the data.
The strata with the highest proportion of failures were the ones with low coverage in the data, which seems reasonable since those would be expected to be the least trustworthy.
The results of the model comparison and investigating the monotonicity assumption under the alternative models can be found at the online supplementary material. 

Figure \ref{fig_sens}$(b)$ shows for each stratum the average boundaries within which the probability $p_{11}$ was constrained among the monotonicity-consistent posterior draws.
The interval widths are mostly around ten percent-units, suggesting that the constraints imposed by the monotonicity assumption were relatively strict and the treatment-dependent confounder should have had substantial effects on the mediator and the outcome in order to make a significant change in the results when varying the sensitivity parameter.

We also investigated the agreement of the step monotonicity assumption with the data. Under the step monotonicity assumption, 3279 ($41\%$) of the posterior draws resulted in an improper joint distribution of $(U_{a^*},U_a)$ for at least one stratum, suggesting the step monotonicity to be an overly restrictive assumption.

\subsection{Total, direct and indirect effects}

\begin{table*}[h!]
\caption{The posterior means and $95\%$ posterior intervals (in parentheses) of the direct ($DE$), indirect ($IE$) and total ($TE$) effects under the two alternative functionals and different choices of the sensitivity parameter $\rho$.\label{table_results}}
\tabcolsep=0pt
\begin{tabular*}{\textwidth}{@{\extracolsep{\fill}}lcccccc@{\extracolsep{\fill}}}
\toprule
 & \multicolumn{3}{@{}c@{}}{Three-year legacy} & \multicolumn{3}{@{}c@{}}{Current change} \\
 
 \cline{2-4} \cline{5-7} \\
$\rho$ & $DE$ & $IE$ & $TE$ & $DE$ & $IE$ & $TE$ \\
\midrule
min & \makecell{$0.530$ \\ $(-1.74, 2.20)$} & \makecell{$0.888$\\$(0.068,1.95)$}&  \makecell{$1.57$ \\ $(-0.082, 2.95)$}
 & \makecell{$1.24$\\$(-0.012,2.74)$} & \makecell{$0.243$\\$(-0.142,0.680)$} &  \makecell{$1.61$ \\ $(0.313, 3.09)$} \\
0.5 & \makecell{$0.600$\\$(-1.72,2.24)$} & \makecell{$0.970$\\$(0.133,2.10)$} &  \makecell{$1.57$ \\ $(-0.082, 2.95)$}
& \makecell{$1.31$\\$(0.051,2.81)$} & \makecell{$0.306$\\$(-0.081,0.748)$} &  \makecell{$1.61$ \\ $(0.313, 3.09)$} \\
max & \makecell{$0.689$\\$(-1.57,2.34)$} & \makecell{$1.05$\\$(0.217,2.13)$} & \makecell{$1.57$ \\ $(-0.082, 2.95)$}
 & \makecell{$1.37$\\$(0.116,2.87)$} & \makecell{$0.369$\\$(-0.018,0.814)$} &  \makecell{$1.61$ \\ $(0.313, 3.09)$}  \\
\bottomrule
\end{tabular*}
\end{table*}

Table \ref{table_results} shows the posterior means and $95\%$ credible intervals (CI) of the direct, indirect and total effects by choosing the stratum specific sensitivity parameters so that the mediational effects were either minimised or maximised. 
In addition, the value $\rho=0.5$ was used homogeneously in all strata.
The estimated effects remained virtually the same regardless of the choice.
We therefore discuss the results with the choice $\rho = 0.5$. 

Under the three-year legacy parameterisation, the estimated indirect treatment effect, i.e. the effect mediated through the change in BMI, amounts to roughly one year of additional time without T2D over the 15 years since the treatment onset ($95\%$ CI 0.13--2.1).
The point estimate of the direct treatment effect was 0.60 years with a very wide credible interval extending to the negative side.
Within the posterior range of the total effect, the indirect effect was almost always positive. 
The smaller the total effect, the more prominent was the role of the indirect effect (see the online supplementary material).
The estimated total treatment effect amounts to 1.6 years of additional time remaining free of T2D ($95\%$ CI $-0.08$--$3.0$ years).
For comparison, a nonparametric estimator of the total effect based on the areas under Kaplan-Meier curves of the two treatment groups yielded an estimate of 1.7 years of additional T2D-free time ($95\%$ confidence interval $0.85$--$2.6$).

The current change parameterisation resulted in a more pronounced direct effect of $1.3$ years of additional time free of T2D ($95\%$ CI $0.05$--$2.8$), while the indirect effect was $0.31$ years ($95\%$ CI $-0.08$--$0.75$).
The total effect remained similar to that under the three-year legacy parameterisation, however, the credible interval became slightly narrower.

As it could be considered a plausible \emph{a priori} assumption, that the effects of the intervention cannot be negative, the results are also shown by discarding posterior draws ($23\%$ of the total MCMC samples under the three-year legacy and $8\%$ under the current change parameterisation) giving a negative estimate for the direct or the indirect effect (Table \ref{table_results_positive}). Under such a restriction, the proportion mediated, i.e. $IE/TE$, becomes a well defined quantity and describes the proportion of the total effect that may be attributed to the indirect mechanism. 
With the three-year legacy parameterisation, the direct and the total effects were larger, while the indirect effect was slightly smaller. The estimated proportion mediated was roughly half, however, and the $95\%$ credible interval extended over the entire $[0,1]$ interval.
With the current change parameterisation there was little change in results compared with the unrestricted analysis as only $8\%$ of the posterior draws were discarded. 
The estimated proportion mediated was $20\%$ with a $95\%$ credible interval from $2\%$ to $64\%$.

\begin{table*}[h!]
\caption{The posterior means and $95\%$ posterior intervals (in parentheses) of the direct ($DE$), indirect ($IE$) and total ($TE$) effects under the two alternative functionals after discarding Markov chain Monte Carlo samples giving negative treatment effect estimates. The sensitivity parameter was set at $\rho = 0.5$.\label{table_results_positive}}
\tabcolsep=0pt
\begin{tabular*}{\textwidth}{@{\extracolsep{\fill}}lccccc@{\extracolsep{\fill}}}
\toprule
Model & $DE$ & $IE$ & $TE$ & \makecell{Proportion \\ mediated} & \makecell{N discarded \\ MCMC samples} \\ 
\midrule
Three-year & \makecell{$0.980$ \\ $(0.076, 2.30)$} & \makecell{$0.831$ \\ $(0.150,1.61)$} & \makecell{$1.81$ \\ $(0.860,3.04)$} & \makecell{$0.487$ \\ $(0.092,0.940)$} & \makecell{1776 ($23\%$)} \\ 
Current & \makecell{$1.33$ \\ $(0.235,2.77)$} & \makecell{$0.331$ \\ $(0.032,0.751)$} & \makecell{$1.66$ \\ $(0.523,3.11)$} & \makecell{$0.229$ \\ $(0.020,0.639)$} & \makecell{628 ($8\%$)} \\ 
\bottomrule
\end{tabular*}
\end{table*}

\subsection{Sensitivity analyses}\label{section_sens}

\begin{figure}[h!]
	\includegraphics[width=\linewidth]{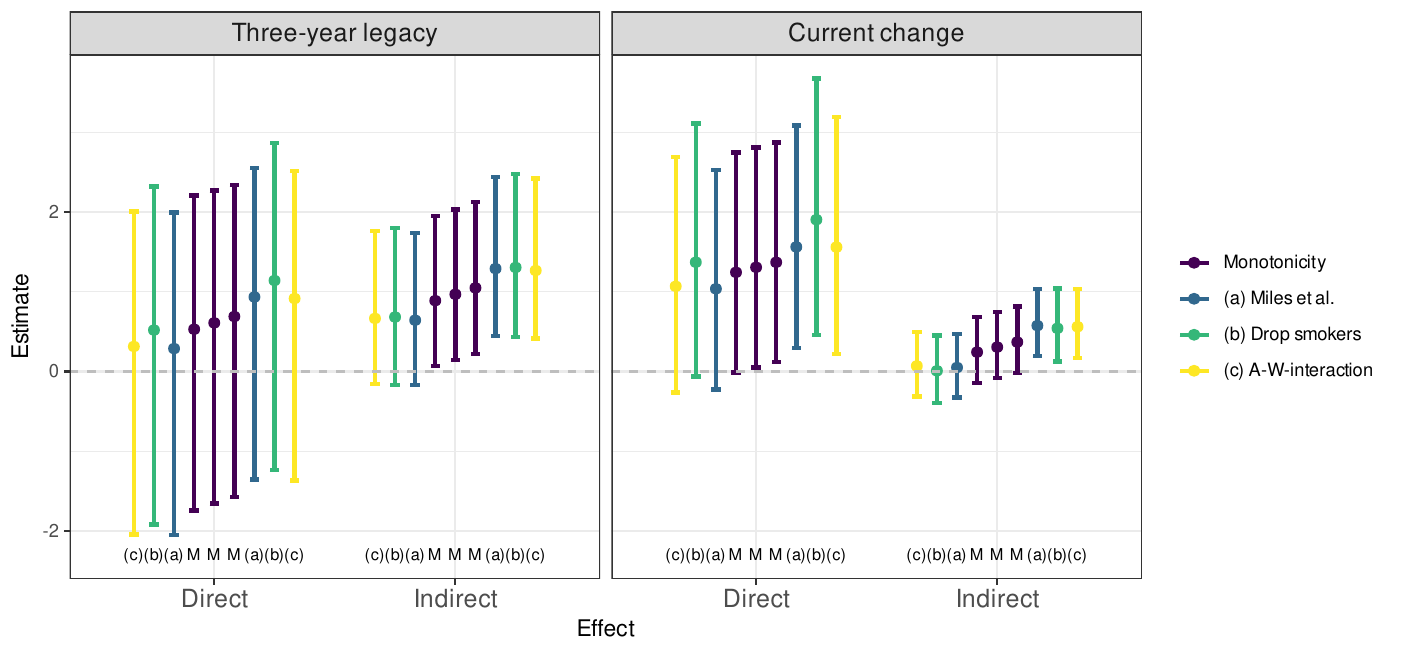}
	\centering
	\caption{The lower and upper bounds for the estimates of direct and indirect effects and their $95\%$ credible intervals under relaxing the monotonicity assumption. The estimates in the middle (with identifier `M') are the results obtained under the monotonicity assumption.
	 To the left and right are the lower and upper bounds obtained using the approach of \citet{Miles2017} with either (a) the full sample; (b) dropping smokers from the sample; or (c) using the treatment-dependent confounder model with all one-way interactions between the treatment and each baseline covariate.}
	\label{fig_sensitivity}
\end{figure}

For sensitivity analyses, we used the approach of \citet{Miles2017} to find lower and upper bounds for the estimates relaxing the monotonicity assumption,  i.e., optimising the expressions with respect to a joint probability matrix $(U_{a^*},U_a)$ without the constraints implied by the monotonicity assumption.
We considered three scenarios: using either (a) the full data and the same model for $U$ as in the primary analyses; (b) the same model for $U$ but removing current smokers as a potential outlier group from the data; or (c) the full data and the model for $U$ with all treatment-covariate interactions included.
The lower and upper bounds for the estimates of the direct and indirect effects and their $95\%$ credible intervals are shown in Figure \ref{fig_sensitivity}, along with the results obtained under the monotonicity assumption.
As all the estimates are very similar, we conclude that the results are robust against violations of the monotonicity assumption.

Interestingly, the estimated direct effects appear to be higher when current smokers are excluded from the data while the indirect effects are largely unchanged.
This suggests that the treatment might be less favourable for smokers than the others.
The total effect estimates using the simple estimator based on areas under Kaplan-Meier curves resulted in a total effect of $1.9$ years for never smokers or former smokers and negative total effect of $-1.3$ years for current smokers.
This comparison, however, is very uncertain as smokers comprised a small group of only 30 individuals.

\section{Discussion}\label{section_discussion}

We addressed causal mediation in the case of a longitudinal mediator and a time-to-event outcome in a randomised lifestyle intervention study, where some of the lifestyle changes acted as treatment-dependent confounders. 
Such situations may arise in interventional studies aimed at reducing the risk for an adverse health outcome through lifestyle changes.
When the interest lies in the treatment effect mediated through changes in a biological risk factor, such as body weight in the current study, other lifestyle changes will act as treatment-dependent confounders, complicating the identification of mediational effects.
The methods described here can be applied to address such situations in order to elucidate mechanisms by which treatments affect health outcomes.

Previously, \citet{Tchetgen2014} showed that if the effect of treatment on a binary-valued treatment-dependent confounder is monotonic, mediational effects can be identified from empirical data.
We here extended their approach to an ordinal trichotomous case, where a single sensitivity parameter (for each baseline covariate stratum) needs to be specified to identify the mediational effects.
The sensitivity parameter determines the joint probability distribution of the two counterfactuals, $(U_{a^*}, U_{a})$, of the treatment-dependent confounder and identifies the component of the total effect that controls how the effect due to the additive interaction of the mediator and the treatment-dependent confounder is divided into the direct and indirect effects.
Our approach is related to partial identification, where bounds for the effect estimates can be determined by optimising the equations \eqref{eq_de_ie} with respect to the unidentified joint distribution without any structural assumptions on the joint distribution of $(U_{a^*},U_{a})$ \citep{Miles2017}.
The monotonicity assumption may be seen as a structural assumption that sets some of the joint probabilities to zero.
This also implies a necessary condition on the marginals of the joint distribution of $(U_{a^*},U_a)$, allowing empirical assessment of the tenability of the monotonicity assumption.

Other approaches to deal with treatment-dependent confounding have been introduced. 
Identification can be retained if the counterfactuals of the treatment-dependent confounder are assumed to be independent or if one is a deterministic function of the other \citep{Robins2010}, or if there is no additive interaction between the treatment and the treatment-dependent confounder \citep{Tchetgen2014}.
Another option would be to switch the estimation target from natural direct and indirect effects to their interventional analogues, which are identifiable even in the presence of treatment-dependent confounders but do not, in general, share the same interpretation as the natural effects \citep{VanderWeele2014b, Nguyen2022,Miles2023}.

We treated the longitudinal mediator (BMI) as a functional entity, i.e. a smooth function describing the underlying trajectory which the mediator follows and of which observations endowed with stochastic deviations were made over time.
The contribution of the stochastic deviations to the association between the mediator and the time-to-event outcome was assumed to be negligible, hence promoting the underlying function itself as the effective mediator.
An implicit assumption was that the structure given for the trajectory function is flexible enough to capture the longitudinal process to a relevant extent.
By defining the mediator as a functional entity, we avoided the practical difficulties dealing with a high-dimensional mediator, as well as the conceptual challenges that might arise from having to control for post-treatment values of the mediator.
Similar approaches treating longitudinally measured mediators as functional entities have been previously investigated in the mediation analysis literature \citep{Lindquist2012,Coffman2023,Zeng2023}.

We chose the restricted survival time as the time-to-event outcome to avoid issues arising from unmeasured heterogeneity under outcomes defined conditionally on previous survival \citep{Aalen2015}.
For instance, if there were an unobserved genetic factor influencing T2D risk, conditioning on survival at any post-treatment time would induce a selection bias, potentially invalidating the analysis.
In addition, as one third of the study participants did not develop T2D, a substantial tail proportion of the event time distribution remained unobserved.
This may introduce the risk of misspecifying the parametric shape of the tail distribution.
To address this, we chose the restriction time as 15 years since the baseline to represent a clinically meaningful time period providing reasonable number of follow-up visits and T2D events.

We used a joint modelling framework to estimate the parametric models.
The association structure was induced by incorporating the latent mediator trajectory in the survival submodel and also specifying a shared random effects structure \citep{Gould2015,Papageorgiou2019}.
As the mediator was considered a latent quantity, individual-specific mediator trajectories needed to be estimated and the uncertainty regarding the estimation appropriately propagated to the estimates of the causal effects.
This propagation of uncertainty was straightforwardly handled in the Bayesian set-up of computations.

From the causal perspective, shared random effects between the longitudinal and survival submodels represent latent confounding between the mediator and the outcome.
The joint modelling framework can thus be employed to account for such unmeasured confounding \citep{Zheng2021}.
We here used the random intercept of the mediator trajectory as a shared random effect to reflect a latent property of the individual that may influence both the mediator trajectory and the time-to-event outcome.
In our empirical application this might be, for example, an unmeasured metabolism-related genetic factor.

The primary goal of our empirical application was to decompose the treatment effect on the restricted mean T2D-free time into indirect (mediated through weight reduction) and direct (all other mechanisms) effects.
Using the three-year legacy parameterisation, the estimated indirect effect translated to one year of additional time without T2D over the 15 years after the start of follow-up, with a $95\%$ credible interval ranging from 1.5 months to two years.
The estimated total effect was 1.6 years with a $95\%$ credible interval from negative one month to positive three years.
These results suggest that the early weight reduction may indeed constitute a major mechanism through which a lifestyle intervention affects long-term T2D-free survival, accounting for two thirds of the total effect as calculated crudely from the point estimates.
However, using the current change parameterisation, the indirect effect was estimated to be considerably smaller (0.31 years, $95\%$ CI $-0.08$--$0.75$ years), accounting, based on the point estimates, roughly $20\%$ of the total effect (1.6 years, $95\%$ CI $0.31$--$3.1$).
Sensitivity analyses relaxing the monotonicity assumption yelded largly consistent results.

Asuming that the treatment effects cannot be negative had little impact on the estimates under the current change parameterisation, and the proportion mediated was estimated to be $23\%$ with the $95\%$ credible interval ranging from $2\%$ to $64\%$.
With the three-year legacy parameterisation, however, $23\%$ of the numerical posterior samples were discarded and the direct effect estimate increased and the indirect effect estimate decreased compared with the unrestricted case.
The decrease of the indirect effect estimate after removing samples with negative effects was due to the fact that the two mediational effects had a strong negative correlation.
Since the posterior distribution of direct effect extended far to the negative side, a large number of samples with a very large negative direct effect, and consequently a very large positive indirect effect, were not admissible.

The choice of the legacy parameterisation was motivated by previous literature demonstrating the persistence of the intervention effects on T2D risk long after the discontinuation of the active treatment and diminishing of the acquired group differences in the clinical risk factors \citep{Wilding2014}.
We used the cumulative change at three years as a time-constant predictor in the survival submodel starting from the onset of treatment.
We justify this by noting that the individual trajectories are determined by the observed baseline covariates and the random effects and can thus be interpreted to exist at the baseline, even though they may be learned only by observing the trajectory unfold over time.
In this context, the weight reduction during the first three years was interpreted as a surrogate for some biological process that responds quickly to the initiation of the lifestyle intervention and the induced behavioural changes.
This underlying biological process was then assumed to manifest through the subsequent weight reduction and be the true causal mechanism linking weight reduction to a decrease in T2D risk.

The direct and indirect effects were identified up to sensitivity parameters controlling the probabilities ($p_{11}$) that the two counterfactual levels of the trichotomous treatment-dependent confounder in each stratum both belong to the middle category (corresponding to an absent treatment effect).
The estimated causal effects remained largely unaffected by the sensitivity parameter.
This was likely due to the fact that the constraint imposed by the monotonicity assumption on $p_{11}$ (see eq. \eqref{eq_p11}) was very stringent.
Furthermore, increasing the probability of an absent treatment effect necessarily also increases the probability ($p_{02}$) of a large treatment effect, simultaneously decreasing the probabilities for intermediate treatment effects ($p_{01}$ and $p_{12}$).
If the treatment-dependent confounder has a monotonic effect on the mediator and the outcome, this may be seen as a self-regulating property, since the highest and lowest effects must always be up- or downweighted together.

Our empirical analysis has some potential limitations.
First, the lifestyle score acting as the treatment-dependent confounder was constructed somewhat crudely by summarising variables describing the targeted lifestyle changes over the first three follow-up visits.
As the treatment-dependent confounder is an important part of the causal mechanism, any inaccuracy in its measurement may distort the resulting causal effect estimates.
Second, the generalisability of the results is limited.
The DPS inclusion criteria selected volunteers who were already overweight, had developed impaired glucose tolerance and were between 40 to 65 years of age at the screening visit but had not yet been diagnosed with T2D.
As such, the DPS cohort represents a selected population, that is, individuals at a high risk of T2D who had managed to avoid the disease until a relatively old age.

In conclusion, we investigated causal mediation in longitudinal intervention studies with a time-to-event outcome in the presence of an ordinal treatment-dependent confounder.
Foremost, we showed that assuming monotonicity of the treatment effect on a trichotomous ordinal treatment-dependent confounder, the direct and indirect effects can be identified up to stratum-specific scalar sensitivity parameters.
The time-to-event outcome was defined as a restricted survival time to avoid issues pertaining to measures conditioning on prior survival.
To overcome challenges with a high-dimensional mediator, we treated the longitudinal mediator as a functional entity and employed a joint modelling framework to control for possible unobserved confounding between the mediator and the outcome.
The methodology was applied to decompose the effect of a lifestyle intervention on restricted T2D-free time into an indirect effect through weight reduction and a direct effect involving other mechanisms.
We found some evidence suggesting the existence of a clinically significant indirect effect through weight reduction, however, the magnitude of the estimated indirect effect depended considerably on the assumed effective form of the mediator.
When using the weight change over the first three years as the mediator, the indirect effect accounted for a large fraction of the total effect.
Conversely, when considering the current weight change since baseline as the mediator, the direct effect was substantially larger than the indirect one.
The results remained similar in sensitivity analyses relaxing the monotonicity, indicating robustness to violations of this assumption.


\bibliographystyle{abbrvnat}

\end{document}


\maketitle

\begin{appendix}

\section{Identification of causal effects}\label{app_id}
The assumed causal DAG (Figure 1 in the main text) implies the following conditional independencies:
\begin{enumerate}
\item $T_{a,u,m} \ind \{A,U,M\} \text{  } | \boldsymbol{W},R_0$
\item $M_{a,u} \ind \{A,U\} \text{  } | \boldsymbol{W}$
\item $U_a \ind A \text{  } | \boldsymbol{W}$
\item $M_{a,u} \ind \{U_a,U_{a*}\} \text{  } | \boldsymbol{W}$
\item $T_{a,u,m} \ind \{U_a, U_{a^*},M_{a^*,u'}\} \text{  } | \boldsymbol{W},R_0$.
\end{enumerate}
In addition, we make the consistency assumption stating that the observed outcome for an individual with given treatment, mediator and treatment-dependent confounder is equivalent to the potential outcome we would have observed, had the individual been assigned those values for the treatment, mediator and intermediate confounder.
This means, for example, that $T^{(i)}_{a,m,u} = (T^{(i)}|A_i=a,M_i=m,U_i=u$).
Missing data (i.e, right censorings of the survival process and drop-outs in the mediator process) are assumed  missing at random.

In what follows, we first derive the expressions for the expectation of the nested counterfactual $\mathbb{E}(T_{a,M_{a^*}})$ in terms of the observed data and the unidentified joint probability of $(U_a,U_{a'})$.
Then, we proceed to show how applying the additive decomposition leads to expressions which may then be arranged into the terms $DE^{(r)}$, $IE^{(r)}$, $\Delta_{DE}$, $\Delta_{IE}$ and $\delta$ (see Eq. (2) in the main text).
To make the notations more succinct, we hereafter omit the baseline confounders $\boldsymbol{W}$ as well as the latent confounder $R_0$.

Invoking the assumptions 1--5 along with the consistency assumption (denoted by `c') and using factorisation of probabilities (denoted by `f'), the expression for $\mathbb{E}(T_{a,M_{a^*}})$ can be obtained as follows:

\begin{equation*}
\begin{aligned}
\mathbb{E}(T_{a,M_{a^*}}) &= \mathbb{E}(T_{a,U_a,M_{a^*,U_{a^*}}}) \\
&\stackrel{\text{(f)}}{=} \int_{t,u,m} tP(T_{a,U_a,M_{a^*,U_{a^*}}}=t | M_{a^*,U_{a^*}}=m,U_a=u)\times \\ & \hspace{.1\linewidth}P(M_{a^*,U_{a^*}}=m,U_a=u) \dd{t}\dd{u}\dd{m} \\
&\stackrel{\text{(c)}}{=} \int_{t,u,m} tP(T_{a,u,m}=t | M_{a^*,U_{a^*}}=m,U_a=u)\times \\ & \hspace{.1\linewidth}P(M_{a^*,U_{a^*}}=m,U_a=u)\dd{t}\dd{u}\dd{m} \\
&\stackrel{\text{(f)}}{=} \int_{t,u,u',m} tP(T_{a,u,m}=t | M_{a^*,U_{a^*}}=m,U_a=u,U_{a^*}=u')\times \\
& \hspace{.2\linewidth} P(M_{a^*,U_{a^*}}=m,U_a=u|U_{a^*}=u') P(U_{a^*}=u') \dd{t}\dd{u}\dd{u'}\dd{m} \\
&\stackrel{\text{(c)}}{=} \int_{t,u,u',m} tP(T_{a,u,m}=t | M_{a^*,u'}=m,U_a=u,U_{a^*}=u')\times \\
&\hspace{.1\linewidth} P(M_{a^*,u'}=m,U_a=u|U_{a^*}=u') P(U_{a^*}=u') \dd{t}\dd{u}\dd{u'}\dd{m} \\
&\stackrel{(5)}{=} \int_{t,u,u',m} tP(T_{a,u,m}=t)P(M_{a^*,u'}=m,U_a=u,U_{a^*}=u')  \dd{t}\dd{u}\dd{u'}\dd{m} \\
&\stackrel{(4)}{=} \int_{t,u,u',m} tP(T_{a,u,m}=t)P(M_{a^*,u'}=m)P(U_a=u,U_{a^*}=u')  \dd{t}\dd{u}\dd{u'}\dd{m} \\
&\stackrel{(1)}{=} \int_{t,u,u',m} tP(T_{a,u,m}=t|A=a,U=u,M=m)P(M_{a^*,u'}=m) \times \\ & \hspace{.1\linewidth}P(U_a=u,U_{a^*}=u')  \dd{t}\dd{u}\dd{u'}\dd{m} \\
&\stackrel{(2)}{=} \int_{t,u,u',m} tP(T_{a,u,m}=t|A=a,U=u,M=m)\times \\
&\qquad P(M_{a^*,u'}=m|A=a^*,U=u')P(U_a=u,U_{a^*}=u')  \dd{t}\dd{u}\dd{u'}\dd{m} \\
&\stackrel{\text{(c)}}{=} \int_{t,u,u',m} tP(T=t|A=a,U=u,M=m)P_M(m|A=a^*,U=u')\times \\ & \hspace{.1\linewidth}P(U_a=u,U_{a^*}=u')  \dd{t}\dd{u}\dd{u'}\dd{m}\\
&= \int_{u,u',m} \mathbb{E}(T|A=a,U=u,M=m)P_M(m|A=a^*,U=u') \times \\ & \hspace{.1\linewidth}P(U_a=u,U_{a^*}=u') \dd{u}\dd{u'}\dd{m}.
\end{aligned}
\end{equation*}
When $a=a^* \equiv a'$, the expression becomes
\begin{equation*}
\begin{aligned}
\mathbb{E}(T_{a',M_{a'}}) &= \int_{u,m} \mathbb{E}(T|A=a',U=u,M=m)P_M(m|A=a',U=u)P(U_{a'}=u)\dd{u}\dd{m} \\
& \stackrel{(\text{f},\text{c},3)}{=}\int_{u,m} \mathbb{E}(T|A=a',U=u,M=m)P_M(m|A=a',U=u)P_U(u|a')\dd{u}\dd{m},
\end{aligned}
\end{equation*}
since we do not need to simultaneously involve the conflicting worlds where $A$ was set to different values. 

Following the derivation given in the supplement of \citet{Tchetgen2014}, by applying the additive decomposition for $\mathbb{E}(T|a,m,u)$ we obtain
\begin{equation*}
\begin{aligned}
\mathbb{E}(T_{a,M_{a^*}}) &=  \int_{u,u',m}\big[\beta_m(a,m) + \beta_u(a,u) + \beta_{m,u}(a,m,u) + \bar\beta_{a}(a) \big]P_M(m|a^*,u') \times \\
& \qquad \qquad P(U_a=u,U_{a^*}=u')\dd{u}\dd{u'}\dd{m} \\ \\
&= \int_{u,u',m}\beta_m(a,m)P_M(m|a^*,u')P(U_a=u,U_{a^*}=u')\dd{u}\dd{u'}\dd{m}\text{ }+ \\ 
& \quad \int_{u,u',m}\beta_u(a,u)P_M(m|a^*,u')P(U_a=u,U_{a^*}=u')\dd{u}\dd{u'}\dd{m}\text{ }+ \\
&\quad \int_{u,u',m} \bar\beta_a(a)P_M(m|a^*,u')P(U_a=u,U_{a^*}=u')\dd{u}\dd{u'}\dd{m}\text{ }+ \\ 
& \quad \int_{u,u',m}\beta_{m,u}(a,m,u)P_M(m|a^*,u')P(U_a = u, U_{a^*}=u')\dd{u}\dd{u'}\dd{m} \\ \\
&= \int_{u',m}\beta_m(a,m)P_M(m|a^*,u')P_U(u'|a^*)\dd{u'}\dd{m}\text{ }+ \\ 
& \quad \int_{u}\beta_u(a,u)P_U(u|a)\dd{u}\text{ }+  \bar\beta_a(a)\text{ }+ \\ 
& \quad \int_{u,u',m}\beta_{m,u}(a,m,u)P_M(m|a^*,u')P(U_a = u, U_{a^*}=u')\dd{u}\dd{u'}\dd{m}.
\end{aligned}
\end{equation*}
Applying the above expressions for $\mathbb{E}(T_{a,M_{a}})$ and $\mathbb{E}(T_{a,M_{a^*}})$, the definitions of the direct and indirect effects (see Eq. (1) in the main text) lead to the following expressions:
\begin{equation*}
\begin{aligned}
DE &= DE^{(r)} - \Delta_{DE} + \delta, \\
IE &= IE^{(r)} + \Delta_{IE} - \delta, \\
DE^{(r)} &=  \int_{m}\big[ \beta_m(a,m) - \beta_m(a^*, m) \big] P_M(m|a^*) \dd{m}\text{ } + \\
& \quad \int_{u}\big[ \beta_u(a,u)P_U(u|a) - \beta_u(a^*,u)P_U(u|a^*) \big]\dd{u}\text{ } + \\
& \quad  \big[ \bar \beta_{a}(a) - \bar \beta_{a}(a^*) \big], \\
IE^{(r)} &= \int_{m}  \beta_m(a,m) \big[ P_M(m|a)-P_M(m|a^*) \big]\dd{m}, \\
\Delta_{DE} &= \int_{m,u'}\beta_{m,u}(a^*,m,u')P_M(m|a^*,u')P_U(u'|a^*) \dd{m}\dd{u'}, \\ 
\Delta_{IE} &=  \int_{m,u}\beta_{m,u}(a,m,u)P_M(m|a,u)P_U(u|a)\dd{m}\dd{u},\\
\delta &= \int_{m,u,u'}\beta_{m,u}(a,m,u)P_M(m|a^*,u')P(U_a=u,U_{a^*}=u') \dd{m}\dd{u}\dd{u'}.\\
\end{aligned}
\end{equation*}

Estimates of the causal effects can be obtained by estimating the components comprising the empirical formula.
In the presence of (non-informative) censoring, the correct estimation of the expectations of the restricted survival time requires that the censoring is dealt with in the estimation step.
Similarly, the drop-out in the longitudinal process is required to be at most missing at random in order to enable unbiased estimation of the mediator trajectories.

\section{Ordinal treatment-dependent confounder with $K$ values}

Assume $(P)$ is a monotonic probability matrix.
Denote $(P)_{jk} = P(U_{a^*}=j, U_a = k | \boldsymbol{W})$ and its marginals as $\mu_k = (P)_{\cdot k}$ and $\phi_j = (P)_{j\cdot}$, $j,k = 1,\dots,K$.
Conditionally on the marginals, there are $\frac{1}{2}(K-1)(K-2)$ free probabilities in $(P)$.
There are $\frac{1}{2}(K-1)(K-2)$ pairs of indices $(l,r)$ such that $1 < l \leq r < K$.
For any such pair, define $\mu^L = \{\mu_1,\dots,\mu_{l-1}\}$, $\mu^C = \{\mu_l,\dots,\mu_r\}$, $\mu^R = \{\mu_{r+1},\dots,\mu_K\}$, and define similarly $\phi^L$, $\phi^C$ and $\phi^R$.
For each $(l,r)$ it then holds that
\begin{equation}\label{eq_conditions}
p_{\min}(l,r) = \max\{ 0, 1 - ||\phi^L|| - ||\mu^R|| \} \leq ||\phi^C \cap \mu^C|| \leq
\min\{ ||\phi^C||, ||\mu^C||\} = p_{\max}(l,r),
\end{equation}
where $||\cdot||$ denotes the sum of the elements (probabilities) contained within the set (see the derivation below).
There are thus $\frac{1}{2}(K-1)(K-2)$ constraints for the same number of independent linear combinations of the free parameters.

Let $\mathbb{P}$ denote the full set of probabilities in $(P)$.
The monotonicity assumption implies that $||\mathbb{P}|| - ||\phi^L \cup \mu^R|| - ||\phi^C \cap \mu^C|| = 0$.
Since $||\phi^L \cap \mu^R|| \geq 0$, the lower limit of \eqref{eq_conditions} follows from that
\begin{equation}\label{eq_proof}
1 - (||\phi^L|| + ||\mu^R||) = ||\mathbb{P}|| - (||\phi^L \cup \mu^R|| +  ||\phi^L \cap \mu^R||) \leq ||\mathbb{P}|| - ||\phi^L \cup \mu^R|| = ||\phi^C \cap \mu^C||,
\end{equation}
where the equality holds if $||\phi^L \cap \mu^R|| = 0$.
The upper limit follows trivially from the fact that $\phi^C \cap \mu^C$ is a subset of both $\phi^C$ and $\mu^C$ and that each element in the sets must be non-negative.
The equality for the upper bound holds if either $||\phi^C \cap \mu^R|| = 0$ or $||\phi^L \cap \mu^C||=0$.
The lower and upper limits can be estimated from data, and bounds for the causal effect estimates can be obtained by minimising and maximising the effects with respect to the joint probability matrix $(P)$ with restrictions \eqref{eq_conditions}.
Notice that \eqref{eq_conditions} is a necessary but not sufficient condition for monotonicity.
Indeed, for any marginals $(\boldsymbol{\phi},\boldsymbol{\mu})$, one can always construct a non-monotonic joint probability matrix as their outer product.




\section{Step monotonicity}\label{app_step}

In this section, we describe a special case of the monotonicity assumption referred to as step monotonicity.
We show that under step monotonicity, the joint probability of the counterfactuals of the treatment-dependent counfounder is identified and its consistency with the observed data can be empirically assessed.
Let $U$ be an ordinal variable taking values $1,\dots,K$.
Assuming step monotonicity assumes that treatment can, on the individual-level, either have no effect on $U$ or bring it one level higher.
The joint distribution of $(U_a,U_{a^*})$ can then be expressed as a $K\times K$ matrix $P$, where $(P)_{jk} = P(U_a=k,U_{a^*}=j) = 0$, if $k\not \in \{j,j+1\}$.

Denote the row totals corresponding to the marginal probabilities $P(U_{a^*}=j)$, as $\phi_j$, $j=,\dots,K$, and the column totals corresponding to the marginal probabilities $P(U_a=k)$ as $\mu_k$, $k=1,\dots,K$. 
Given the marginals, the probability parameters of the joint distribution are identified, as there are $2K-1$ probability parameters and $2K-1$ restrictions imposed by the marginals.
The joint probability matrix with $N=5$, for example, would correspond to the following $5\times 5$ table
\begin{center}
\begin{tabular}{c c|c c c c c | c}
& & & &$U_a$ & & & \\
& & 1 & 2 & 3 & 4 & 5 & \\
\hline
& 5 & 0 & 0 & 0 & 0 & $p_{55}$ & $\phi_5$ \\
& 4 & 0 & 0 & 0 & $p_{44}$ & $p_{45}$ & $\phi_4$ \\
$U_{a^*}$ & 3 & 0 & 0 & $p_{33}$ & $p_{34}$ & 0 & $\phi_3$ \\
& 2 & 0 & $p_{22}$ & $p_{23}$ & 0 & 0 & $\phi_2$ \\
& 1 & $p_{11}$ & $p_{12}$ & 0 & 0 & 0& $\phi_1$ \\
\hline
& & $\mu_1$ & $\mu_2$ & $\mu_3$ & $\mu_4$ & $\mu_5$ & 1
\end{tabular}
\end{center}
from which it is straightforward to see that if the marginals are known, the elements can be obtained as
$$
p_{ii} = 1 - \sum_{k=1}^{i-1} \phi_k - \sum_{h=i+1}^K \mu_h, \quad p_{i(i+1)} = 1 - \sum_{k=i+1}^K \phi_k - \sum_{h=1}^i \mu_h,
$$
The data are consistent with the step monotonicity assumption if the elements $p_{jk}$ form a proper joint distribution, i.e., all the elements are between $0$ and $1$.

\section{Alternative models for treatment-dependent confounder}

To assess whether a more flexible parameterisation of the model for the treatment-dependent confounder should be preferred, we also considered models with interaction between the treatment and the baseline covariates. 
The models were compared using Pareto smoothed importance sampling leave-one-out cross validation \citep{Vehtari2015}.
The differences in the expected log pointwise predictive masses and their standard errors are shown in Table \ref{table_loo}.
As there was no indication in favour of any of the more flexible parameterisations, the model with no interaction effects (i.e. model (7) of the main text) was chosen.

\begin{table}[h!]
\centering
\caption{Results of the model comparison. The column `ELPPM difference' refers to the difference in the expected log pointwise predictive mass as compared to the best performing model (i.e., one with interaction between treatment and sex) and `SD' to the standard deviation of the difference. The model `none' refers to the simplest model containing no interactions and is here preferred since none of the more complex models could be shown to outperform it. \label{table_loo}}
\tabcolsep=0pt
\begin{tabular*}{.65\textwidth}{@{\extracolsep{\fill}}lcc@{\extracolsep{\fill}}}
\toprule
\makecell[l]{Interactions\\with treatment} & ELPPM difference & SD \\ 
\midrule
$sex$ & 0 & 0 \\
$none$ & -0.1 & 2.1 \\
$age$ & -1.8 & 3.6 \\
\makecell[l]{\emph{score at baseline}} & -3.5 & 2.7 \\
$smoke$ & -4.0 & 2.4 \\
$all$ & -7.9 & 3.7 \\
\bottomrule
\end{tabular*}
\end{table}

\begin{figure}[h!]
\centering
\includegraphics[width=\textwidth]{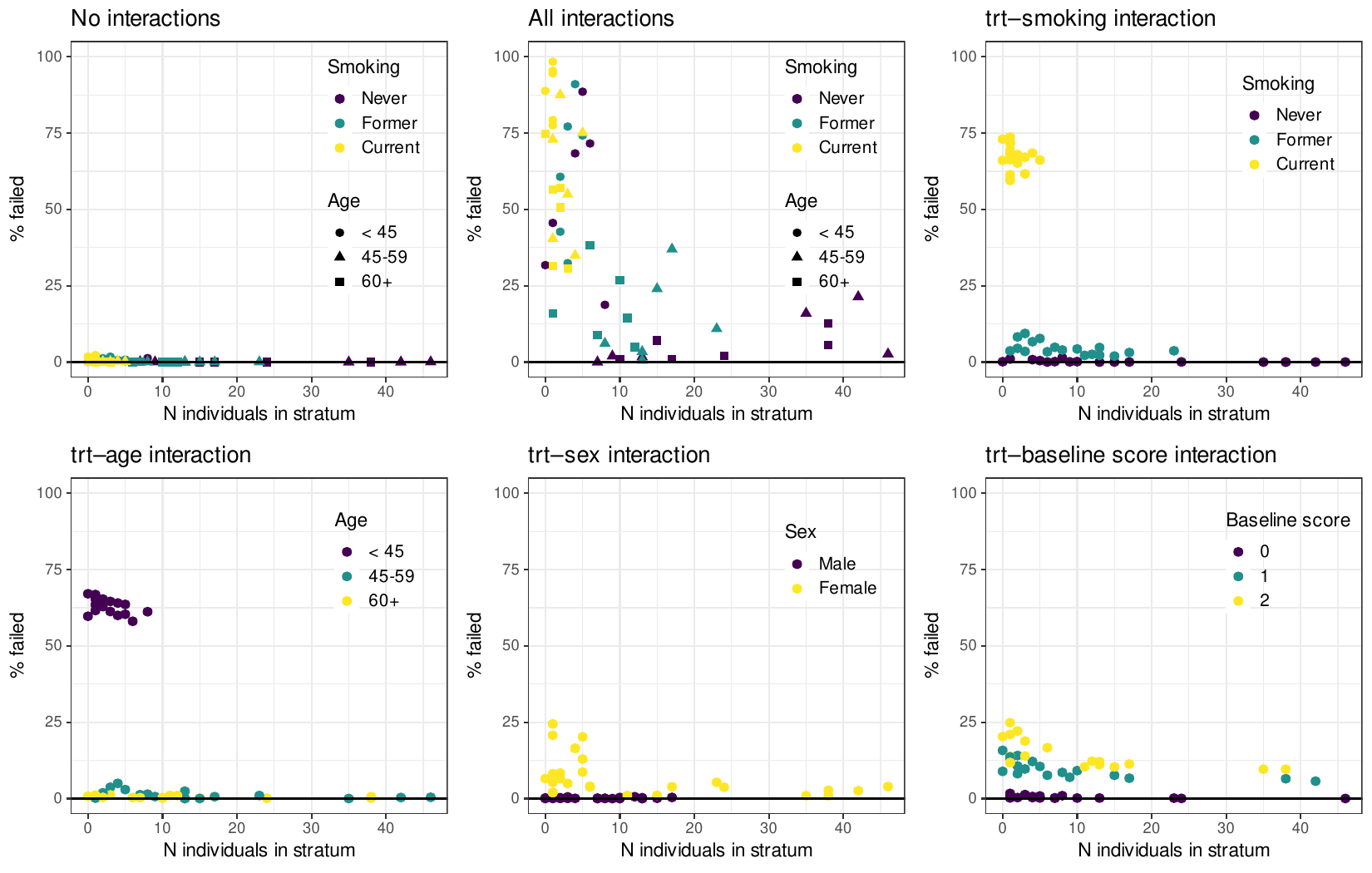} 
\caption{Proportions of posterior samples failing the monotonicity condition under the considered models for the treatment-dependent confounder.}\label{fig_mono_sens}
\end{figure}

The proportions of posterior samples failing the monotonicity assumption (see Eq. (3) in the main text) in each of the compared models are shown in Figure \ref{fig_mono_sens}. 
As there were quite many strata with a large proportion of posterior samples failing the monotonicity condition under the model with all interactions, we chose to conduct a sensitivity analysis estimating lower and upper bounds for the mediational effects under this model and relaxing the monotonicity assumption \citep{Miles2017}.
In addition, as current smokers appeared to prominently feature in the strata challenging the monotonicity assumption, we also performed a sensitivity analysis dropping current smokers.

\section{Joint posterior distributions of the causal effects}\label{app_jointpost}

\begin{figure}[h!]
    \centering
    \includegraphics[width=\textwidth]{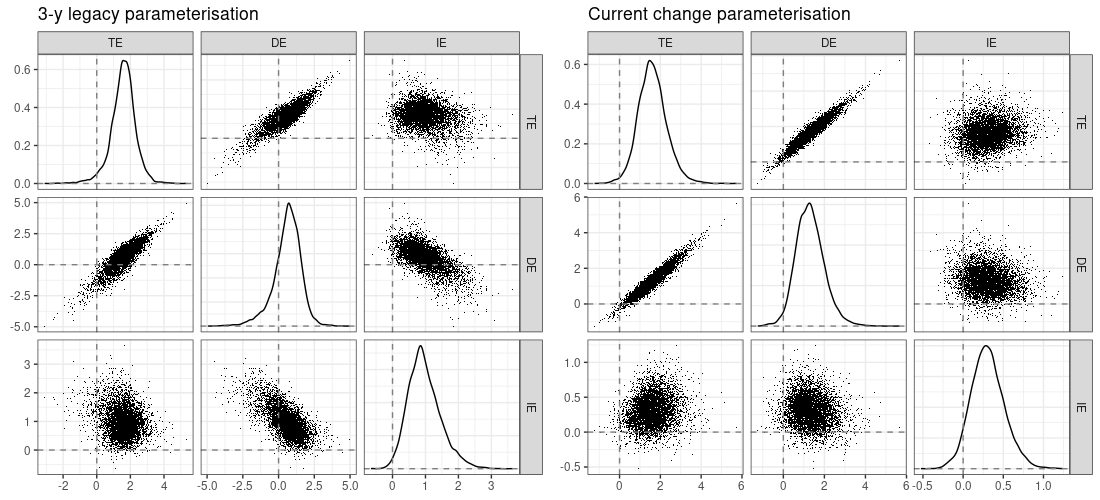} 
    \caption{The pairwise joint posterior distributions of the total (TE), direct (DE) and indirect (IE) effects under the three-year legacy and the current change parameterisations.}\label{fig_app_joint}
\end{figure}

The joint posterior distributions of the total, direct and indirect effects are shown in Figure \ref{fig_app_joint}.
Under the three-year legacy parameterisation, the direct and indirect effects have a strong negative correlation while under the current change parameterisation no appreciable correlation is observed.
The total effect is strongly correlated with the direct effect, representing the fact that the variance of the direct effect is much larger than the variance of the indirect effect.

\end{appendix}